\documentclass[preprint2]{aastex}

\usepackage{epstopdf}

\newcommand{\ob}{$\Omega_{\mathrm{b}}$}
\newcommand{\obh}{$\Omega_{\mathrm{b}}{\cdot}h^2$}

\newcommand{\deu}{D}
\newcommand{\tro}{$^3$He}
\newcommand{\qua}{$^4$He}
\newcommand{\six}{$^{6}$Li}
\newcommand{\sep}{$^{7}$Li}
\newcommand{\be}{$^{7}$Be}
\newcommand{\neu}{$^{9}$Be}
\newcommand{\dix}{$^{10}$B}
\newcommand{\onz}{$^{11}$B}
\newcommand{\hli}{$^4$He, D, $^3$He and $^{7}$Li}
\newcommand{\sbbn}{Standard Big-Bang Nucleosynthesis}

\newcommand{\lap}{\mathrel{ \rlap{\raise.5ex\hbox{$<$}}
	            {\lower.5ex\hbox{$\sim$}}  } }

\shorttitle{BBN to CNO}
\shortauthors{Coc \etal.}

\begin{document}

\title{Standard Big-Bang Nucleosynthesis up to CNO with an improved extended nuclear network}

\author{Alain Coc}
\affil{Centre de Spectrom\'etrie Nucl\'eaire et de Spectrom\'etrie de
Masse (CSNSM), CNRS/IN2P3, Universit\'e Paris Sud, UMR~8609,
B\^atiment 104, F--91405 Orsay Campus, France}
 \author{St\'ephane Goriely and Yi Xu}
 \affil{Institut d'Astronomie et d'Astrophysique, Universit\'e Libre de Bruxelles,
CP 226, Boulevard du Triomphe, B-1050 Bruxelles, Belgium}

\author{Matthias Saimpert and Elisabeth Vangioni}
\affil{Institut d'Astrophysique de Paris, UMR 7095 CNRS, Universit\'e Pierre et Marie Curie, 98 bis Boulevard Arago,
Paris 75014, France}

\begin{abstract}

Primordial or Big Bang nucleosynthesis (BBN)  is one of the three strong evidences
for the Big-Bang model together with the expansion of the Universe
 and the Cosmic Microwave Background radiation.
 In this study, we improve the standard BBN calculations taking into account
 new nuclear physics analyses  and we enlarge the nuclear network until Sodium. 
 This is, in particular,  important to evaluate the primitive value of CNO mass fraction that could affect Population III stellar evolution.
 For the first time we list the complete network of more than 400 reactions with references to the origin of the rates, including $\approx$270 reaction rates 
 calculated using the TALYS code.
 Together with the cosmological light elements, 
  we calculate the primordial Beryllium, Boron, Carbon, Nitrogen and Oxygen nuclei.
   We performed a sensitivity study to identify the important reactions  for  CNO, \neu\ and Boron nucleosynthesis. 
  We reevaluated those important reaction rates using experimental data and/or theoretical evaluations.
 The results are compared with precedent calculations: a primordial Beryllium abundance
 increase by a factor of 4 compared to its previous evaluation, but we note a stability for B/H and for the
 CNO/H abundance ratio that remains close to its previous value of  $0.7\times10^{-15}$.
 On the other hand, the extension of the nuclear network has not changed the \sep\ value, so its abundance is still  3--4 times greater 
 than its observed spectroscopic value. 
\end{abstract}


\today

\section{Introduction}

There are presently three observational evidences for the Big-Bang Model:
the universal expansion, the Cosmic Microwave Background (CMB) radiation
and Primordial or Big-Bang Nucleosynthesis
(BBN).
The third evidence for a hot Big-Bang comes from the primordial
abundances of the ``light elements'': \hli. They are produced during the
first $\approx$20 min of the Universe when it was dense and hot enough for
nuclear reactions to take place.

The number of free parameters entering in 
Standard BBN has decreased with time. The number
of light neutrino families is known from the measurement of the $Z^0$
width by LEP experiments at CERN: $N_\nu$ = 2.9840$\pm$0.0082~\citep{LEP}.
The lifetime of the neutron entering in weak reaction rate
calculations and many nuclear reaction rates have been measured
in nuclear physics laboratories.
The last parameter to have been independently determined is the baryonic density
of the Universe which is now deduced from the observations of the anisotropies
of the CMB radiation coming from the WMAP satellite.
 The number of baryons per photon  which
remains constant during the expansion,  $\eta$, is directly related to \ob\ by
\obh=3.65$\times10^7\eta$ . The WMAP7 gives now :
\obh=0.02249$\pm$0.00056  and $\eta$ = 6.16 $\pm$ 0.15 $\times 10^{-10}$ \citep{WMAP}.
In this context, primordial nucleosynthesis is a parameter free theory and is
the earliest probe of the Universe. We note an overall
agreement except for the \sep. In the literature many studies have been devoted to this lithium problem
\citep[and references therein]{Ang05, Coc07, Cyburtetal2008, Cha11} .

Deuterium, a very fragile isotope, is destroyed after BBN.
Its most primitive abundance is determined from the observation of
clouds at high redshift, on the line of sight of distant quasars. Very few observations of these
cosmological clouds  are available \citep[and references therein]{pettini08}
and the adopted primordial D abundance is given by the average value:

 D/H~$ = (2.82^{+0.20}_{-0.19})\times 10^{-5}$.

After BBN, \qua\ is still produced by stars. Its primitive abundance is deduced from observations
in HII (ionized hydrogen) regions of compact blue galaxies. Galaxies are thought to be formed
by the agglomeration of such dwarf galaxies which are hence considered as more primitive.
The primordial \qua\ abundance $Y_p$ (mass fraction) is given by the extrapolation to zero
metallicity but is affected by systematic uncertainties \citep{aver10, isot10} such as plasma temperature
or stellar absorption. These  most recent determinations based on almost the same set of
observations lead to:

$Y_p = 0.2561 \pm 0.0108$.

Contrary to \qua\ ,  \tro\ is both produced and destroyed in stars so that the evolution of its
abundance as a function is subject to large uncertainties
 and has only been observed in our Galaxy \citep{Ban02},

\tro\ /H~$ = 1.1 \pm 0.2 \times 10^{-5}$.

 \noindent Consequently, the baryometric status of \tro\ is not firmly established \citep{vang03}.

Primordial lithium abundance is deduced from observations of low metallicity stars in the
halo of our Galaxy where the lithium abundance is almost independent of metallicity, displaying
a plateau, the so-called Spite plateau \citep{Spite82}.
This interpretation assumes that lithium has not been depleted at the surface of these stars, so
that the presently observed abundance is supposed to be equal to the initial one. The small
scatter of values around the ÒSpite plateauÓ is an indication that depletion may not have been
very effective.

Astronomical observations of these metal poor halo stars \citep{Ryanetal2000}
 have led to a relative primordial abundance of:

  Li/H~$ = (1.23^{+0.34}_{-0.16}) \times 10^{-10}$ .

\noindent A more recent analysis by \citet{sbordone10} gives:
   
    Li/H~$ =  (1.58 \pm 0.31) \times 10^{-10}$.

\noindent  More generally, \citet{Spite10} have reviewed the last Li observations and
 their different astrophysical aspects. See also \citet{frebel11} for a wide review.

In 2006, high-resolution observations of Li absorption lines in some
very old halo stars have also claimed evidence for a large primitive
abundance  of the weakly-bound isotope $^6$Li \citep{Asp06}. The
\six/\sep\ ratios of $\sim 5\times 10^{-2}$ were found to be about
three orders of magnitude larger than the BBN-calculated value of
\six/\sep$\sim 10^{-5}$.
The key BBN \six\ production mechanism is the
D($\alpha$,$\gamma$)$^6$Li reaction at energies in the range of
50~keV $\le E_{cm} \le$ 400~keV~\citep{Ser04}.
This reaction has very recently been re-investigated \citep{Ham10} confirming
the previous result that \sbbn\ cannot produce \six\ at the required level.
Concerning the \six\ observational status, more recently, however, \citet{Cay07} and \citet{Ste10} have
pointed out that line asymmetries similar to those created by a
$^6$Li blend could also be produced by convective Doppler shifts in
stellar atmospheres. In this context,  these observations have to be confirmed. More detailed analyses
are necessary to firmly conclude about the detection of the \six\ abundance at this level in these metal poor stars
 \citep[see][for a review]{Spite10}.

 We consider in detail in this present study the other isotopes potentially
produced by the \sbbn\ including \neu, \dix, \onz\ and the
CNO isotopes.
The production of \neu, \dix\ and \onz\ (BeB) and CNO isotopes have been studied in the context of standard and
inhomogeneous BBN \citep{Tho93, Tho94, kajino1, kajino2, kajino3, Ioc07, Ioc09}.   
The most relevant analysis concerning CNO in BBN comes from  \citet{Ioc07} who included more than 100 nuclear reactions and predicted 
a CNO/H abundance ratio  of approximately $6 \times 10^{-16}$, with an upper limit of  $10^{-10}$. 
These evaluations had been also performed to provide the initial conditions for the evolution of population III stars.

BeB nucleosynthesis is an important chapter of
nuclear astrophysics. Specifically, rare and fragile nuclei,
they are not generated in the normal
course of stellar nucleosynthesis 
 and are, in fact, destroyed in stellar interiors. This characteristic is reflected by the
low abundance observations of these light  species \citep[and references therein]{prim10, boesg10}.
A glance to the abundance curve  suffices to capture the essence of
the problem: a gap separates He and C. At the bottom of this
precipice rests the trio Li-Be-B. At very low metallicity, the BeB abundance is
less than $10^{-12}$ relatively to hydrogen.
Indeed, they are characterized by the simplicity of their nuclear structure (6
to 11 nucleons) and their scarcity in the Solar System and in stars.
In fact, they
are fragile because a selection principle at the nuclear level
has operated in nature. 
Due to the fact that nuclei with mass  5 and 8  are unstable, BBN has almost stopped at A = 7, 
while nuclear burning in stars bypasses them through the triple--alpha reaction.

After BBN, the formation agents of LiBeB are Cosmic Rays  interacting with
interstellar or circumstellar CNO. Other possible origins have been also
identified, for example supernova neutrino spallation (for \sep\ and \onz\ ). In contrast, \six, \neu\ and \dix\
  are pure cosmic-ray spallative products.  (For a review see \citet{vang00}.)

  Recently, the LiBeB production has been considered in a cosmological context \citep{rollinde06, rollinde08}.
 The non-thermal evolution with redshift of \six, Be, and B in the first structures of the Universe has been studied.
 In this context  cosmic rays are impinging alpha particles and CNO produced by the first massive stars, the so-called Population III stars.
 The computation has been  performed in the framework of hierarchical structure formation and
 reliable \six\ and BeB initial abundances coming from BBN are required to optimize the initial conditions.

Even though the direct detection of primordial CNO isotopes
seems highly unlikely with the present observational techniques at high redshift, 
it is also important to better estimate their \sbbn\ production.
Hydrogen burning in the first generation of stars (Pop III stars) proceeds through the
slow pp chains until enough carbon is produced (through the triple-alpha reaction)
to activate the CNO cycle. The minimum value of the initial CNO mass fraction that
would affect Pop III stellar evolution is estimated to be 10$^{-10}$ \citep{Cas93}  or
even as low as  10$^{-12}$ in mass fraction for the less massive ones \citep{Eks08}.
This is only two orders of magnitude above the  \sbbn\ CNO yield using the current
nuclear reaction rate evaluations of \citet{Ioc07}.

In addition, it has been shown that Pop III stars evolution is sensitive to the triple-alpha
$^{12}$C producing reaction and can be used to constrain the possible variation of the
fundamental constants \citep{Eks10}. This reaction is sensitive to the position of the Hoyle
state, which in turn is sensitive to the values of the fundamental constants. The amount
of produced CNO ($^{12}$C) could affect the HR diagram (CNO versus pp H-burning) and
the final production of $^{12}$C and $^{16}$O in Pop III stars. Hence, it is important to
quantify the amount of primordial CNO present at their birth. In the same context of the
variations of the fundamental constants, $^8$Be (which decays to two alpha particles
within $\sim10^{-16}$~s) could become stable if these constants were only slightly
different. At BBN time, this would possibly allow to bridge the "A=8 gap" and produce
excess CNO. To determine how significant would be this excess, one needs to know
the standard  BBN production of the CNO elements.

Another motivation for this study is the above mentioned dichotomy
concerning the Li abundance.
 At WMAP baryonic density,
\sep\ is produced as $^7$Be that later decays. Nuclear ways to destroy this $^7$Be
have been explored. An increased $^7$Be(d,p)$2\alpha$ cross section has been
proposed by \citet{Coc04} but was not confirmed by experiment described in  \citet{Ang05} unless
a new resonance is present \citep{Cyb09} with very peculiar properties.
Other $^7$Be destruction channels have recently been proposed by \citet{Cha11}
awaiting experimental investigation. Another scenario would be to take advantage
of an increased late time neutron abundance. This is exactly what happens (in the
context of varying constants)  when the $^1$H(n,$\gamma)^2$H rate is decreased.
The neutron late time abundance is increased (with no effect on \qua) so that more
$^7$Be is destroyed by  $^7$Be(n,p)$^7$Li(p,$\alpha)\alpha$.  \citep[see in] [Fig. 1]{Coc07}.

The main difficulty in BBN calculations up to CNO is the extensive network needed,
including n-, p-, $\alpha$-, but also d-, t- and  \tro-induced reactions. Most of the corresponding
cross sections cannot be extracted from experimental data only. This is especially
true for radioactive tritium-induced reactions, or for those involving radioactive targets like
e.g. $^{10}$Be. For some reactions, experimental data, including spectroscopic
data of the compound nuclei, are just inexistent. Hence, for many reactions,
one has to rely on theory to estimate the reaction rates. Previous studies
\citep{Tho93,Tho94,Ioc07} have performed unpublished analyses of experimental data 
but have also apparently extensively used the prescription of
\citet{Fow64} and \citet{Wag67} to estimate many rates. 
These prescriptions often assume a constant astrophysical $S$-factor. This obviously cannot
 be a good approximation for most of the reactions considered in the BBN context. 
 In this study we use at first more reliable rate estimates provided by the TALYS reaction code \citep{TALYS}, next we perform a sensitivity study and, finally improve the rate estimates of the
 most important reactions by dedicated evaluations.

A detailed analysis of all reaction rates and associated uncertainties would be
desirable but is unpractical for a network of $\approx$400 reactions.
So, in Section 2 we present the extended network and the standard thermonuclear reaction rates
used.
In Section 3, we study the sensitivity of the calculated primordial abundances
up to CNO to variation of the rates by a factor of up to 1000 and re-evaluate selected reaction rates.
In Section 4, we present the BBN calculation
 results  and we conclude in section 5. Note that in the annex we give the full list of reactions with 
 the references at the origin of the reaction rates.

\section{Nuclear cross section network}

In this study, we have included 59 nuclides from neutron to $^{23}$Na, linked
by 391 reactions involving n, p, d, t and \tro\ induced reactions and 33 $\beta$-decay processes. Reaction rates
were taken primarily from \citet{NACRE,Des04,ILCCF10b,NACRE2} and other evaluations
when available. The complete list of reactions with associated references to the origin of the rates can 
be found in Table~\ref{t:network}. 
Except for a few (historical) cases,  the "direct reactions", listed in Table~\ref{t:network}, are chosen to have positive Q-value. 
In our code, each of these reactions is systematically supplemented by 
the reverse reaction calculated according to the usual detailed balance prescription \citep{CF88,NACRE}.
In comparison with previous works, the present study includes two specific features, namely the introduction of the new  evaluation of experimental reaction rates (NACRE 2) for target nuclei with $A<16$ and the extensive library of rates for experimentally unknown reactions, including all possible light particle captures. The latter library is based on the Hauser-Feshbach calculation with the TALYS reaction code \citep{TALYS} and though, {\it a priori}, the reaction model is not well suited for the description of the reaction mechanism on such light species, it appears to provide rather fair rates that can be used as a first guess for a sensitivity analysis, as detailed and discussed below.

\subsection{NACRE 2}

In the present paper, use is made of the updated NACRE 2 reaction rate evaluation.
This new evaluation includes new rates for 15 charged-particle transfer reactions and for 19 capture reactions on stable targets with mass number A$<$16. Compared to NACRE \citep{NACRE}, NACRE 2 features in particular (i) the addition to the collected NACRE experimental data of all the post-NACRE ones published in refereed journals till 2010; (ii) the extrapolation of astrophysical $S$-factors to very low energies based on theoretical models, namely the  distorted wave Born approximation (DWBA) for transfer reactions and the potential model for radiative captures \citep{NACRE2b}.  These models are simple albeit trustworthy  for the considered reactions most likely dominated by a direct (rather than a compound) contribution. The experimental data available (usually well above the energy region of astrophysical interest)  are fitted by spline interpolations. If narrow resonances happen to contribute, they are approximated by the  single-level Breit-Wigner formula with varying particle widths.  The Hauser-Feshbach statistical model is used to extrapolate the rates to very high temperatures (i.e. higher relevant astrophysical energies). In this case, the calculations are made with the TALYS  code \citep{TALYS}.

For each individual reaction, recommended as well as upper and lower limits of the reaction rates are given. Uncertainties in the reaction rates are obtained by modifying the optimal model parameters and still allowing for acceptable fits to the experimental reaction data.  All details can be found in \citet{NACRE2b}. The improved theoretical treatment compared to NACRE makes the rates, as well as their respective uncertainty estimates,  more reliable, especially at low temperatures.

\subsection{The TALYS code}
\label{s:talys}

The neutron, proton, deuterium, tritium,$^3$He and $\alpha$-particle capture cross sections are, in a first approximation, estimated with the TALYS nuclear reaction code \citep{Koning08,TALYS}  when not available experimentally or in any other existing compilation of reaction rates. For targets and energies of interest in the present work, TALYS essentially takes the compound mechanisms into account to estimate the total reaction probability, as well as the competition between the various open channels. The cross sections are estimated up to the relevant energies and out of it the Maxwellian-averaged reaction rates for the thermalized target. The calculation includes in the entrance as well as exit channels all single particles of interest here (neutron, proton, deuterium, tritium,$^3$He and $\alpha$-particle). In the exit channel, multi-particle emission is also taken into account. All the experimental information on nuclear masses, deformation, and low-lying states spectra (including $\gamma$-ray intensities) is considered, whenever available. If not, global nuclear level formulas, $\gamma$-ray strength functions, and nucleon and $\alpha$-particle optical model potentials are considered to determine  the excitation level scheme and the photon and particle transmission coefficients. 

Based on the statistical model of Hauser-Feshbach, TALYS is known to be a well-adapted code for medium-mass and heavy target nuclei. Such a model makes the fundamental assumption that the capture process takes place with the intermediary formation of a compound nucleus in thermodynamic equilibrium. The energy of the incident particle is then shared more or less uniformly by all the nucleons before releasing the energy by particle emission or $\gamma$-de-excitation.  The formation of a compound nucleus is usually justified by assuming that the level density in the compound nucleus at the projectile incident energy is large enough to ensure an average statistical continuum superposition of available resonances. However, when the number of available states in the compound system is relatively small, as for light targets, the validity of the Hauser-Feshbach predictions has to be questioned. In this case, other approaches, such a the radiative direct capture within the potential model, need to be considered.

\begin{figure}[h]
\plotone{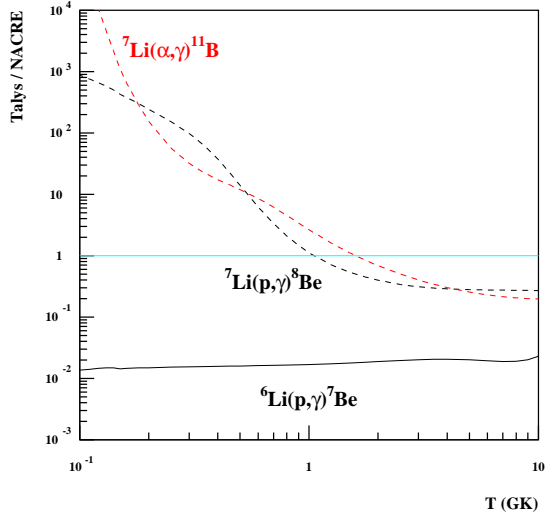}
\caption{Comparison between TALYS   and NACRE rates \citep{NACRE}.}
\label{f:nacrew}
\end{figure}

In the present work, the TALYS estimates are considered as a first guess for the reaction rates on light nuclei in order to test the sensitivity of the abundance calculations with respect to changes in the nuclear reaction rates. As shown in Figs.~\ref{f:nacrew}-\ref{f:ilccf10we}, the agreement between TALYS and experimental rates are surprisingly good in most of the cases, although most of these reactions should not be physically described by the compound nucleus reaction mechanism. In some of these cases, e.g. $^9$Be(p,$\alpha$)$^6$Li, a few resonances are available in the compound nucleus and the predictions are rather satisfactory. In other cases,  $^6$Li(p,$\gamma$)$^7$Be, no resonances is available and the model fails by about 3 orders of magnitude.

In contrast, large deviations can be found between TALYS rates and those estimated  by \citet{Tho93,Tho94}, as shown in Figs.~\ref{f:tho93w}-\ref{f:tho93we}. 
Even the temperature dependence appears to be quite different\footnote{For the  $^9$Li(d,n)$^{10}$Be reaction, this is most probably
due to a typo in \citet{Tho93}  as possibly corrected in the dashed curve in Fig.~\ref{f:tho93w}.}.
In \citet{Tho94}, in several cases, the {\em same rate is used} for the 
(p,n), (p,$\alpha$) and (p,$\gamma$) reaction channels on a given target: an obvious source of errors that can affect the predictions by a few orders of magnitude.

As summarized by the 36 cases studied in Figs.~\ref{f:nacrew}-\ref{f:ilccf10we}, TALYS can globally be expected to provide predictions within 3 orders of magnitude
in the temperature range of interest here. Hence variations of these theoretical rates by three orders of magnitude can in a first step be used in our sensitivity analysis for the BBN abundance calculation. A detailed description and study will then be limited to the reactions affecting the abundance predictions in a significant way.

\begin{figure}[h]
\plotone{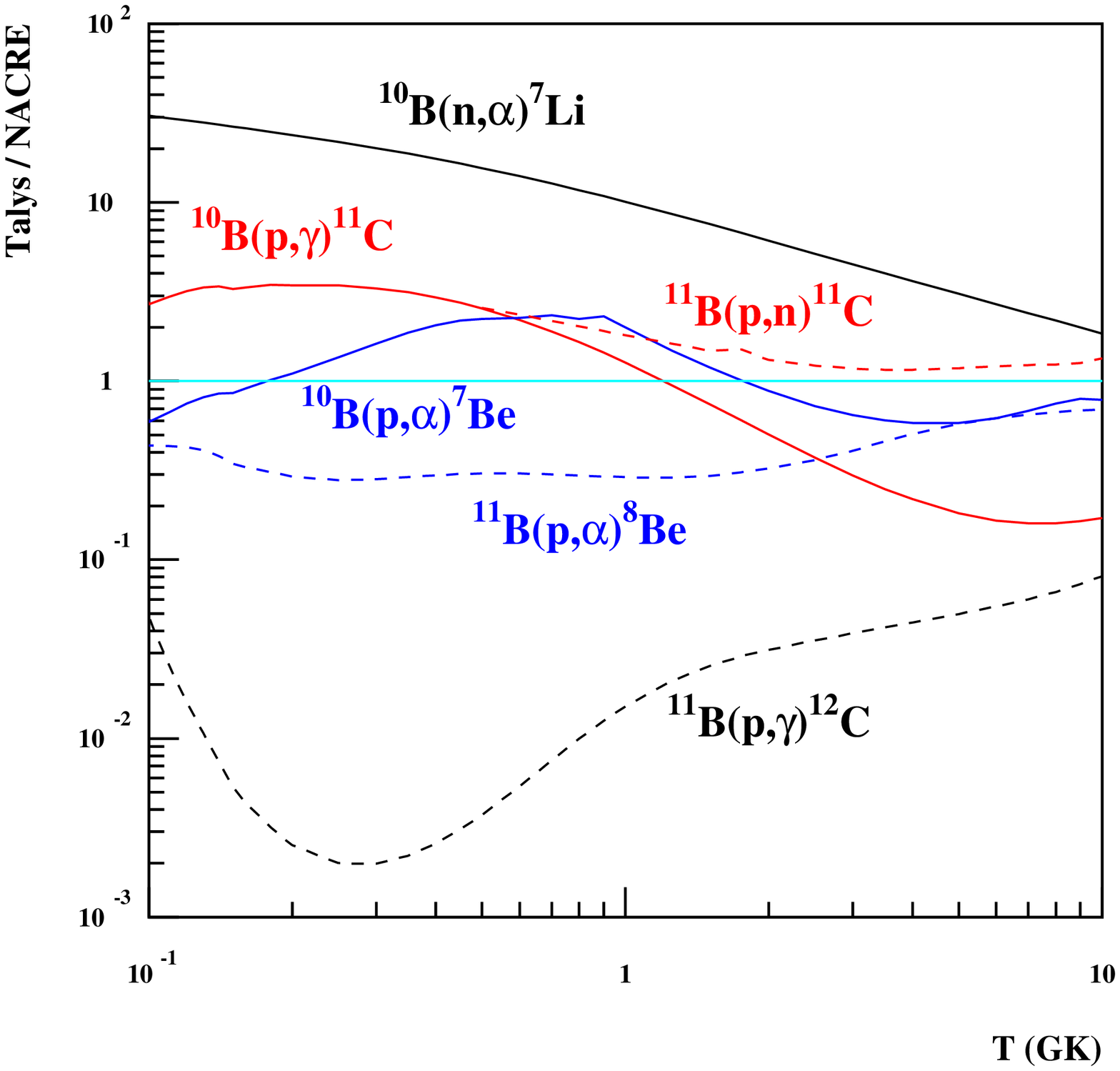}
\caption{Same as Fig.~\ref{f:nacrew}}
\end{figure}

\begin{figure}[h]
\plotone{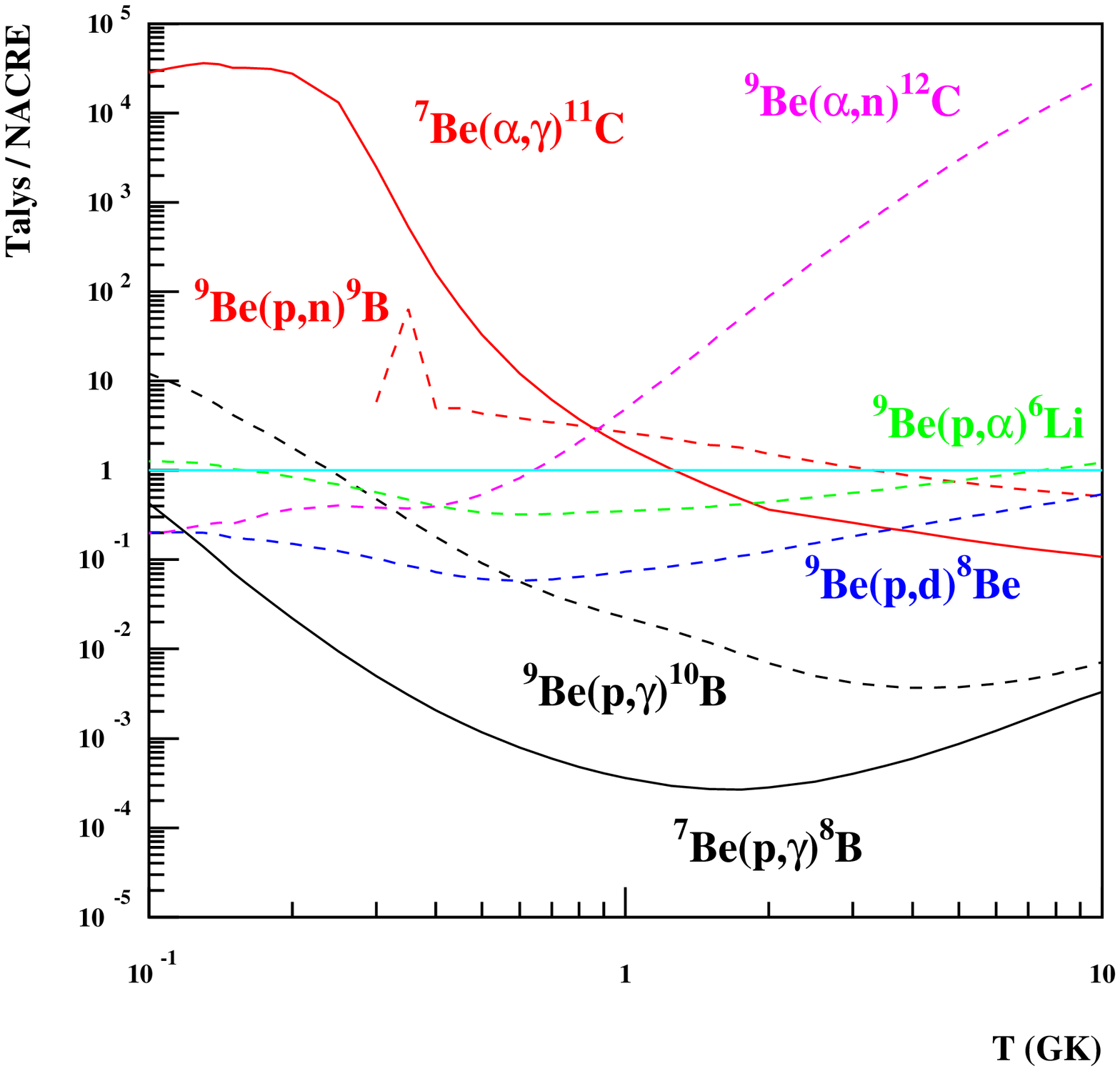}
\caption{Same as Fig.~\ref{f:nacrew}}
\end{figure}

\begin{figure}[h]
\plotone{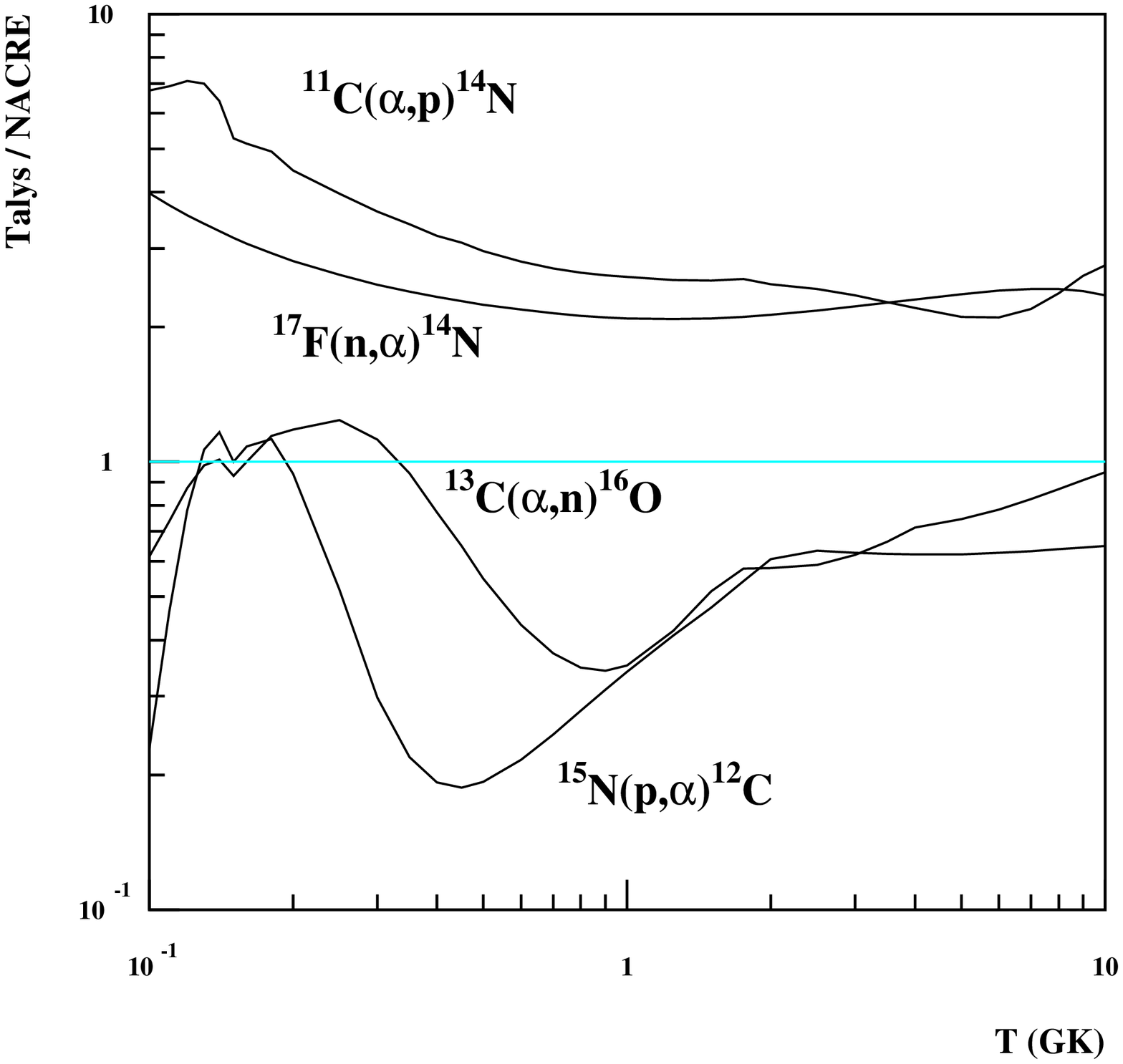}
\caption{Same as Fig.~\ref{f:nacrew}}
\end{figure}

\begin{figure}[h]
\plotone{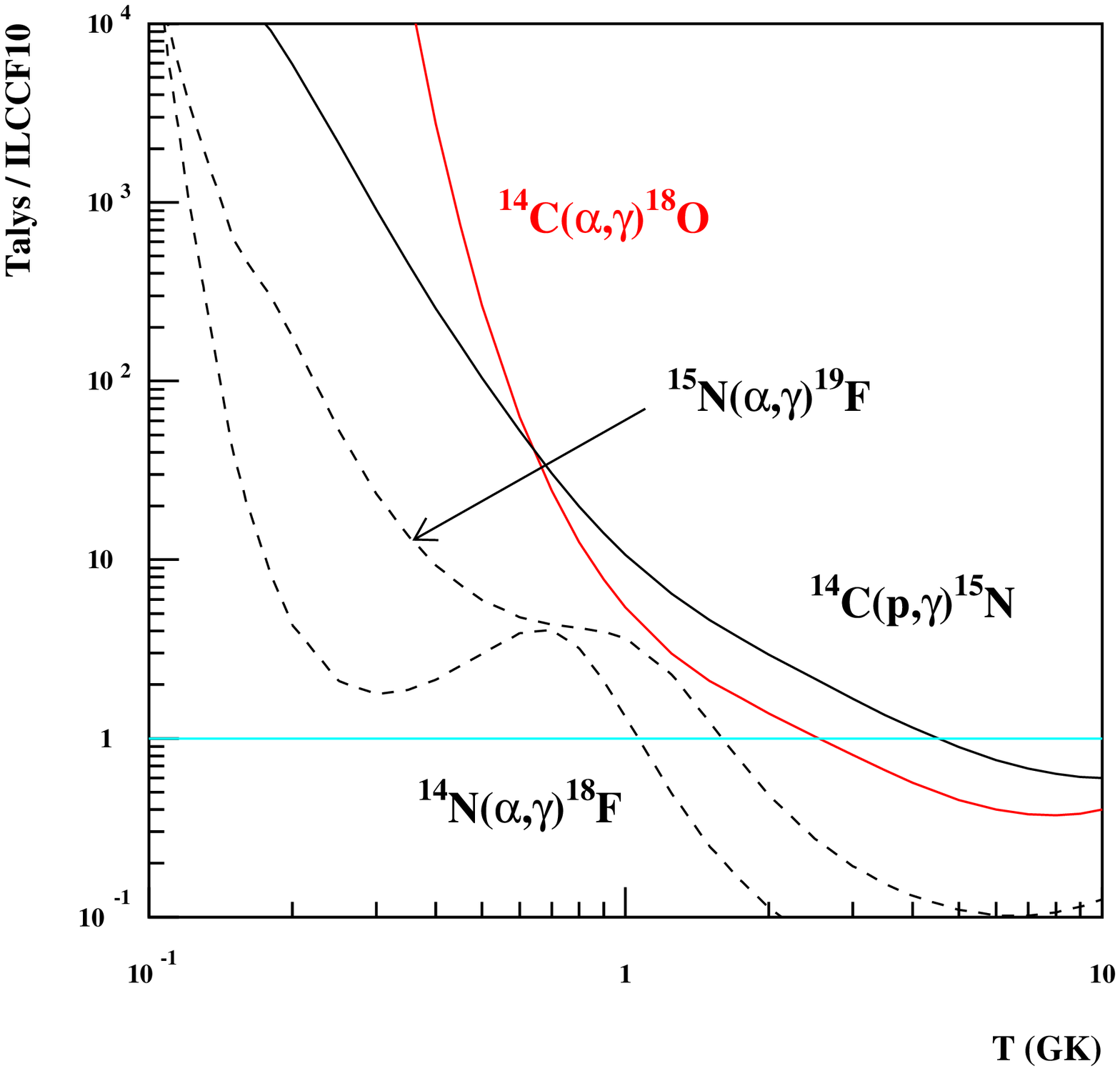}
\caption{Comparison between TALYS  and \citet{ILCCF10b}.}
\label{f:ilccf10w}
\end{figure}

\begin{figure}[h]
\plotone{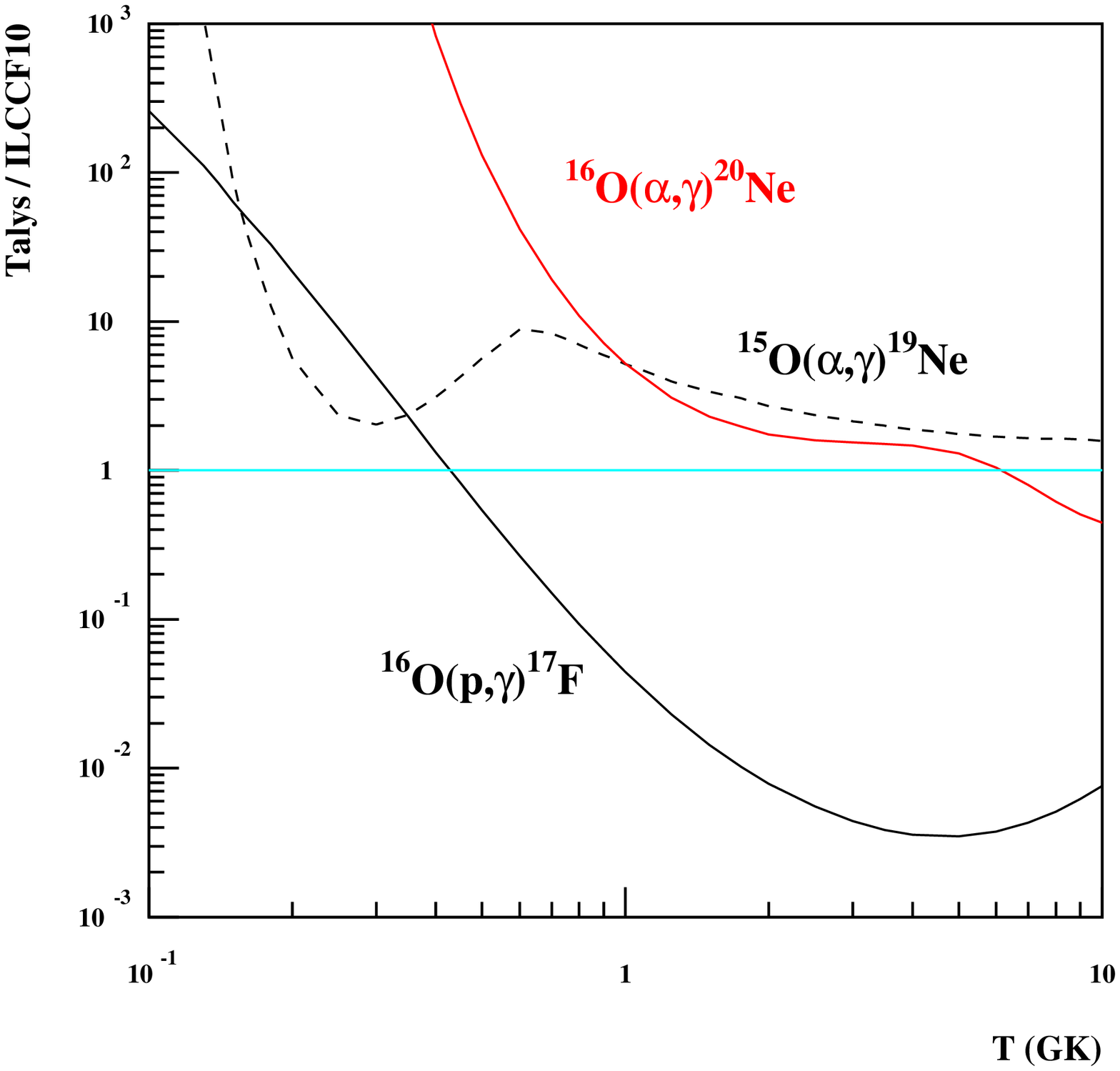}
\caption{Same as Fig.~\ref{f:ilccf10w}}
\end{figure}

\begin{figure}[h]
\plotone{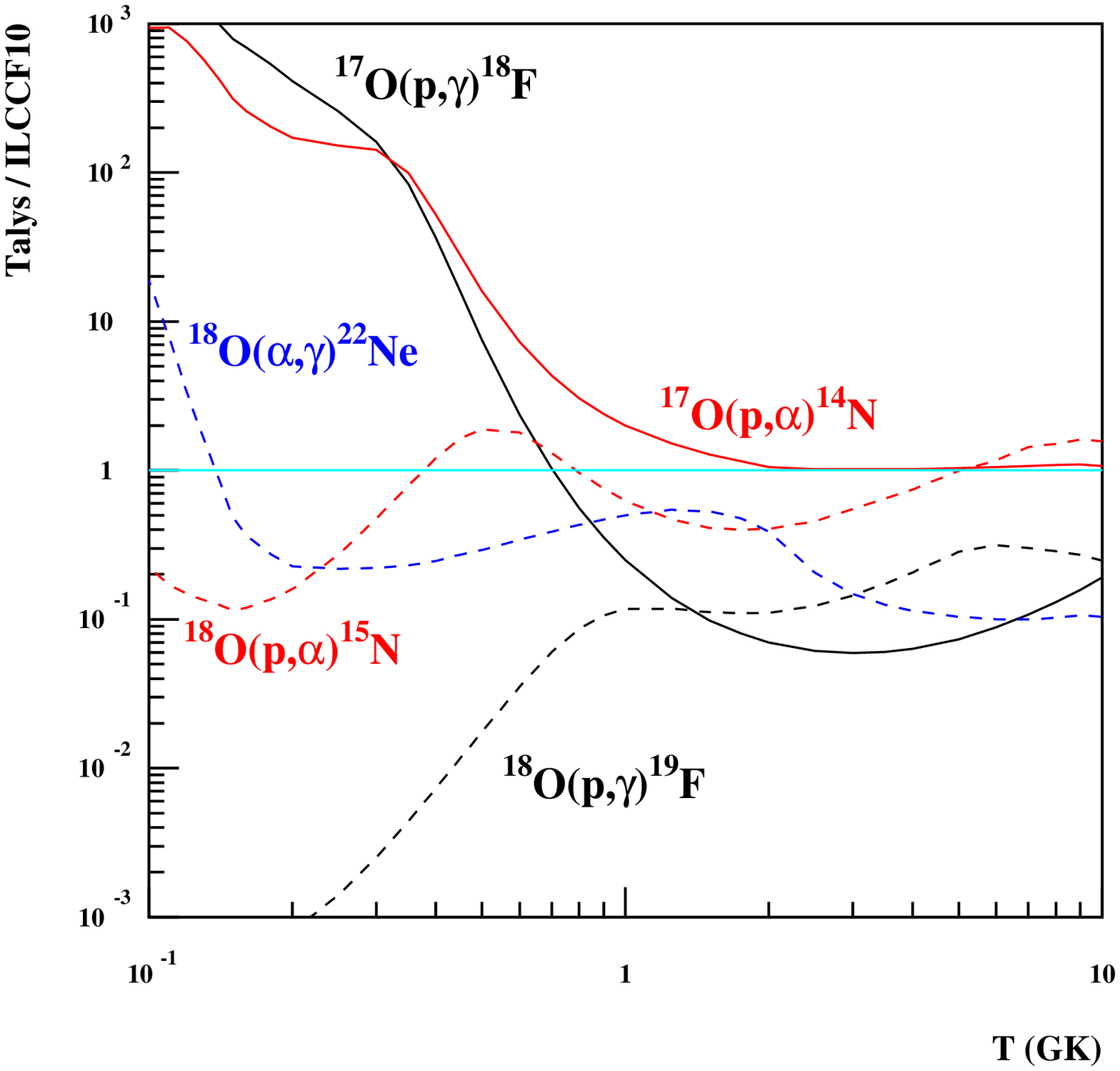}
\caption{Same as Fig.~\ref{f:ilccf10w}}
\end{figure}

\begin{figure}[h]
\plotone{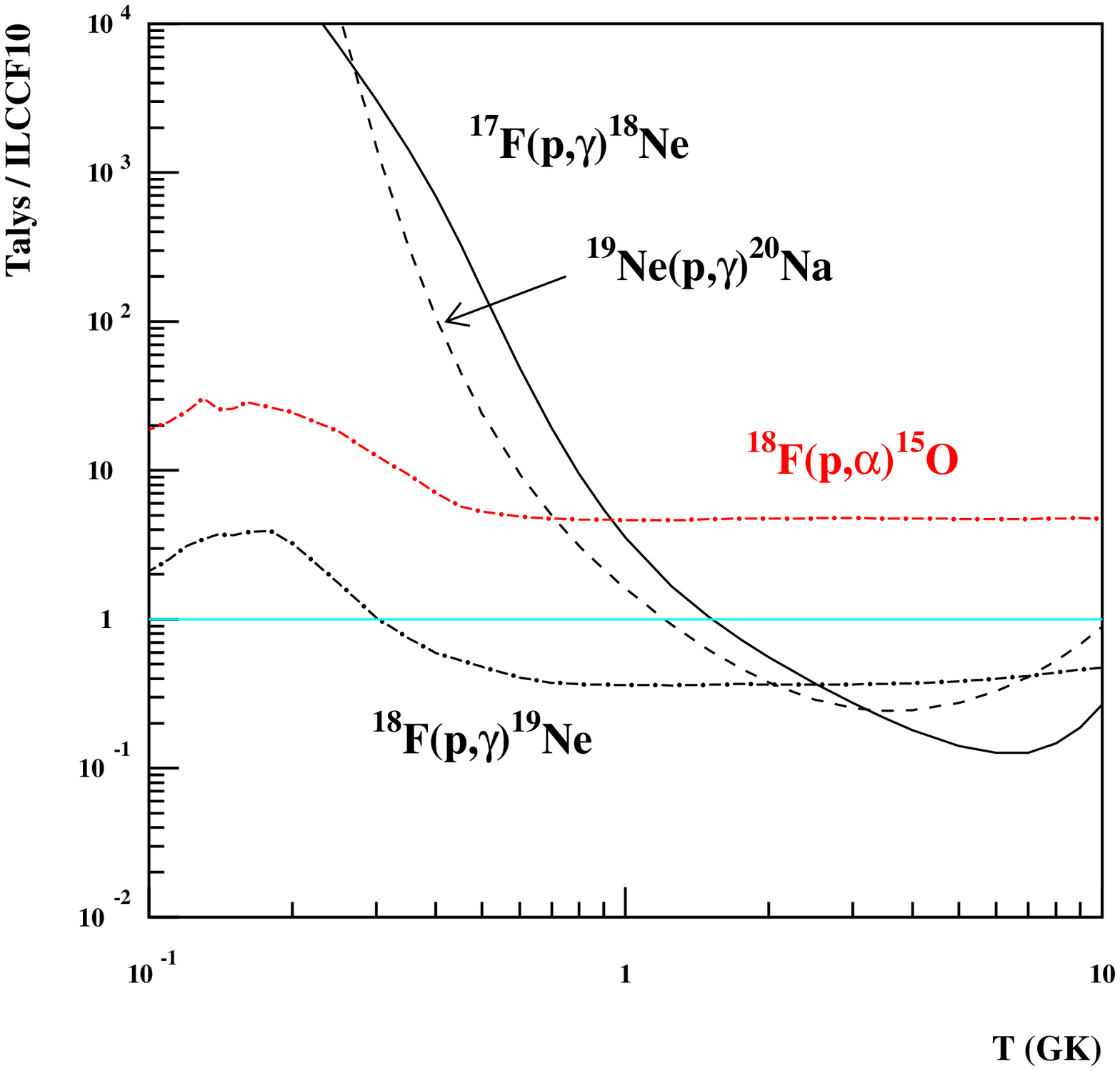}
\caption{Same as Fig.~\ref{f:ilccf10w}}
\label{f:ilccf10we}
\end{figure}

\begin{figure}[h]
\plotone{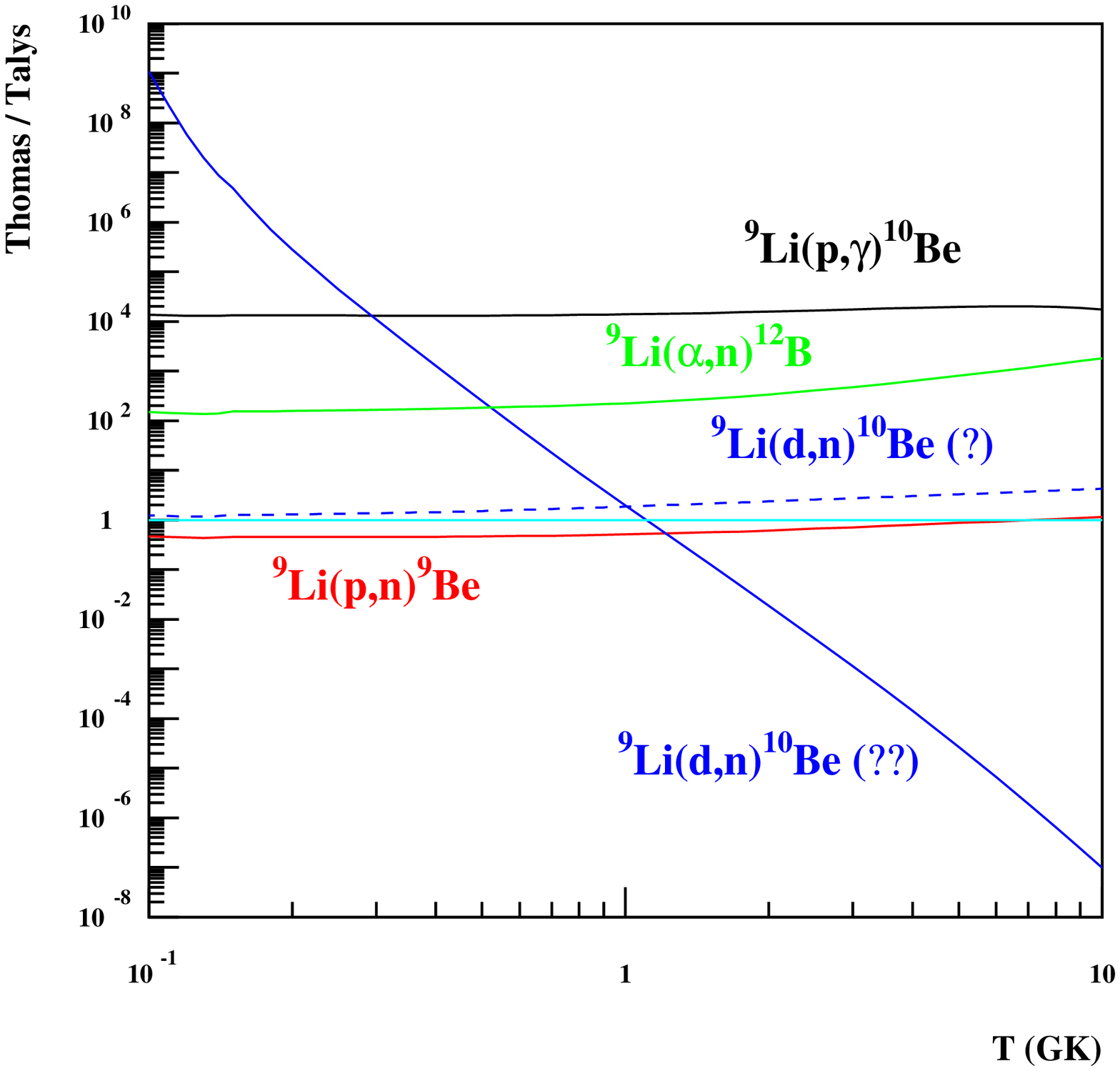}
\caption{Comparison between TALYS  and \citet{Tho93,Tho94}}
\label{f:tho93w}
\end{figure}

\begin{figure}[h]
\plotone{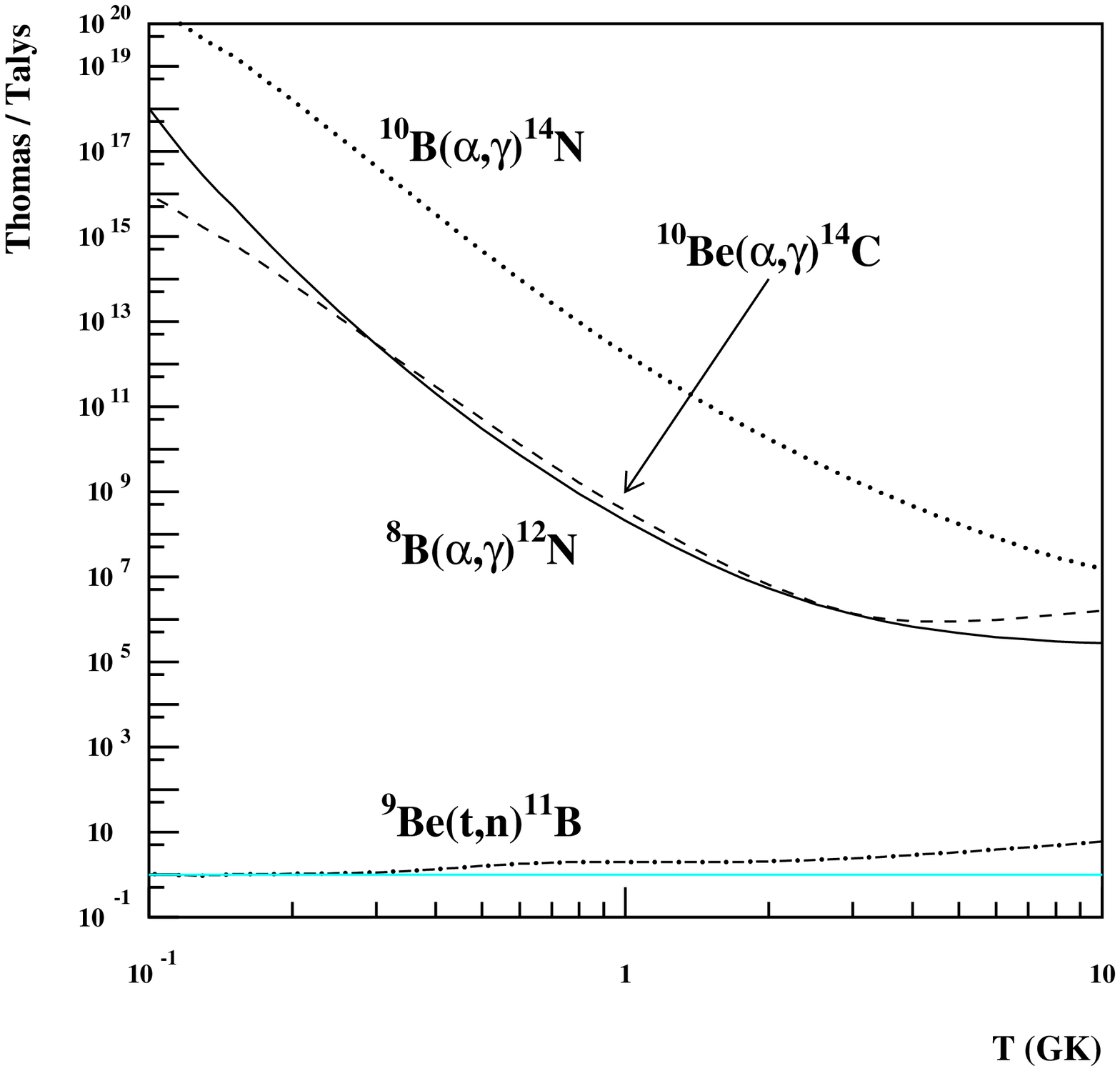}
\caption{Same as Fig.~\ref{f:tho93w}}
\label{f:tho93beb}
\end{figure}

\begin{figure}[h]
\plotone{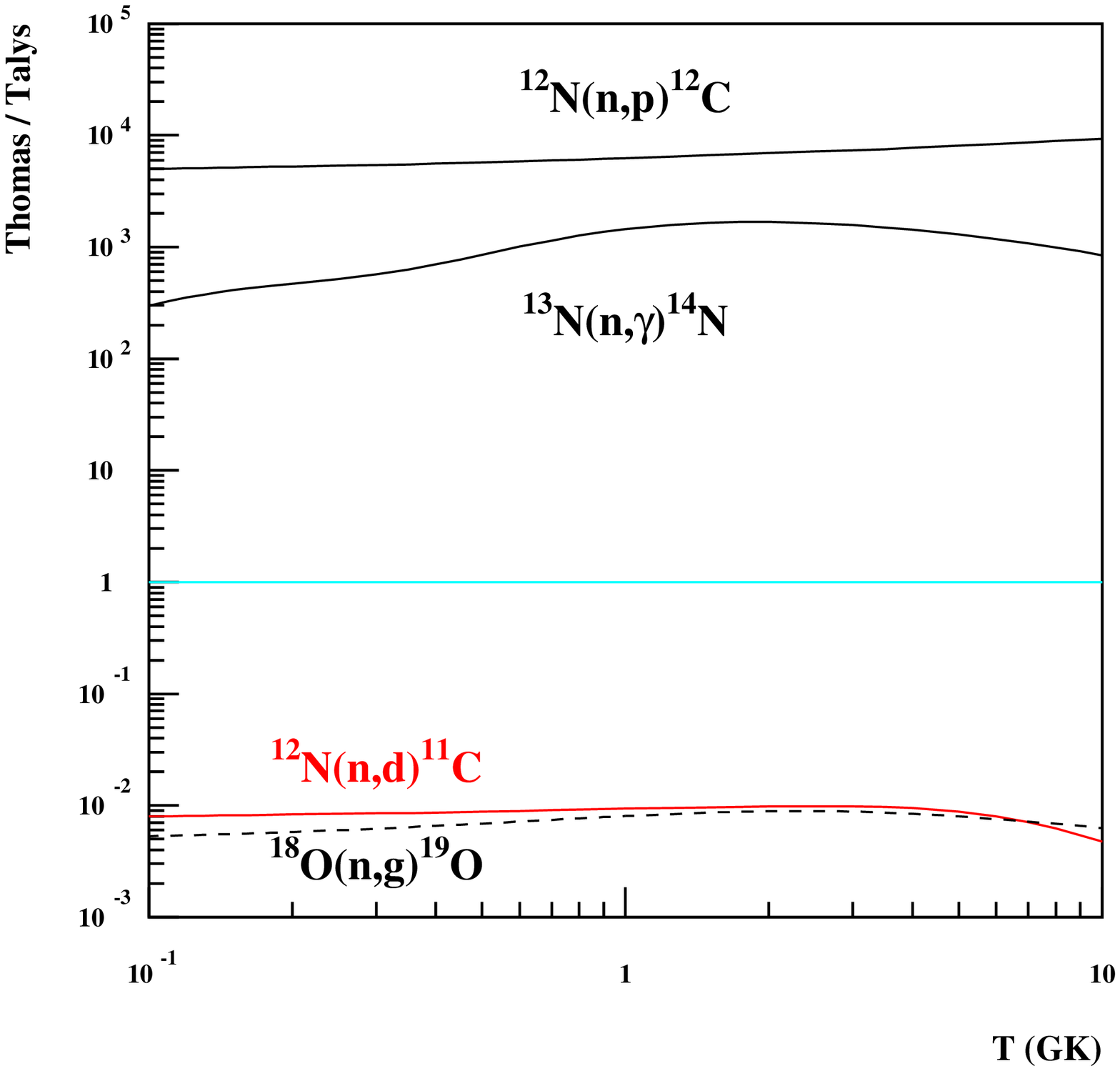}
\caption{Same as Fig.~\ref{f:tho93w}}
\label{f:tho93we}
\end{figure}

\clearpage

\begin{figure}[h]
\plotone{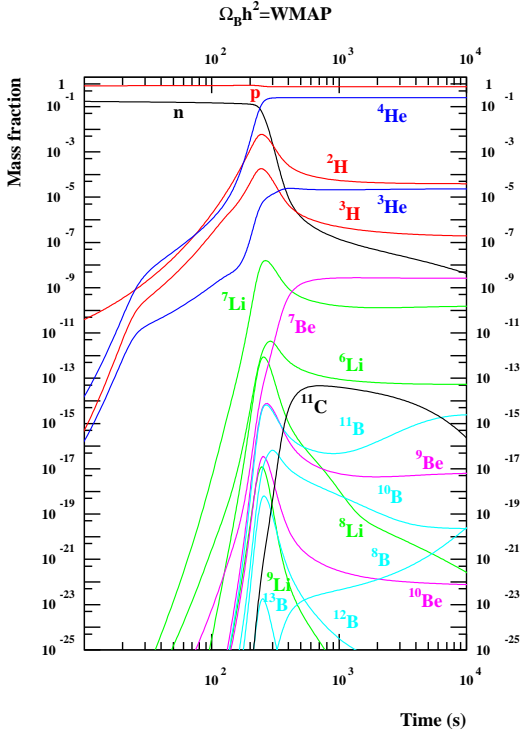}
\caption{(Color online) \sbbn\ production of H, He, Li, Be and B isotopes as a function of time, for the baryon density taken from WMAP7.}
\label{f:cnotime1}
\end{figure}

\begin{figure}[h]
\plotone{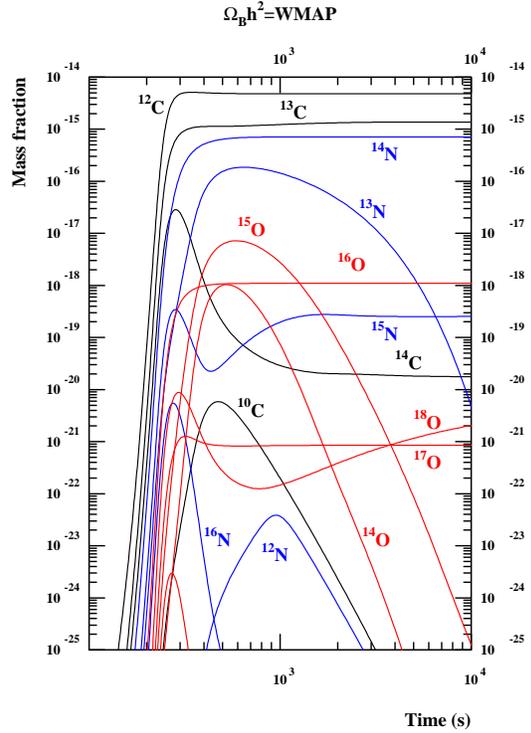}
\caption{(Color online) \sbbn\ production of C, N and O isotopes as a function of time. (Note the different time and abundance ranges
compared to Fig.~\ref{f:cnotime1}.)}
\label{f:cnotime2}
\end{figure}

\begin{figure}[h]
\plotone{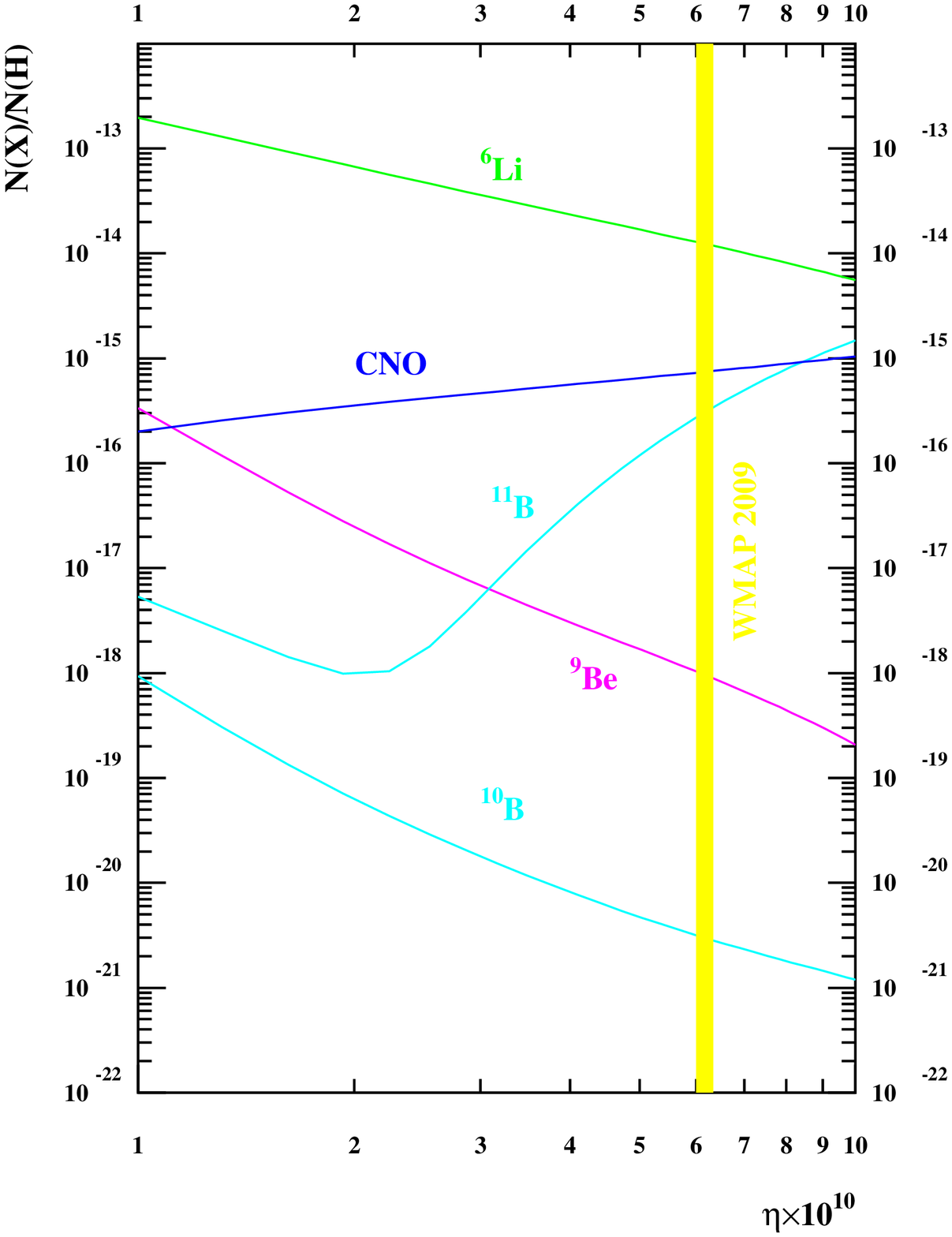}
\caption{(Color online) \sbbn\ production of  \six, Be, B and CNO isotopes as a function of baryonic density.}
\label{f:bebcno}
\end{figure}

\section{Important reactions for \sbbn\ up to CNO}

\subsection{Sensitivity study}

Sensitivity studies have already been performed for the 12 main \sbbn\ reactions
\citep{Nol00,Ser04,Cyb08,CV10} and the many others involved in \hli\ production \citep{Coc04}.
Here, we will consider the impact on \neu, \dix, \onz\ and C, N and O isotopes.

In Table~\ref{t:network}, we list all reactions included in our network. Our sensitivity study
excludes the reactions whose impacts have already been studied in previous work and
whose rate uncertainties are documented in \citet{NACRE,Des04,ILCCF10b,NACRE2}. 
Beta decays
are also excluded because all lifetimes are precisely known \citep{Aud03}.
The sensitivity study is performed for each of the 271 reactions whose rates are obtained
from TALYS and for those whose uncertainties have not been estimated.
Based on the comparison of
Section~\ref{s:talys} between TALYS calculated rates on the one hand and experimental
rates on the other hand we consider at most three orders of magnitude variations around the standard
reaction rates.
More precisely,  to estimate the impact of the reaction rates uncertainties on \sbbn\,
we perform six calculations by changing for each reaction its rate by factors of 0.001, 0.01, 0.1, 10, 100
and 1000 and calculate the relative change in isotopic abundances.  Table~\ref{t:sensib},
displays, for each reaction, isotopes for which the relative changes are larger than 20\%.
Mass fractions of isotopes with A$\geq$12 are added together into CNO.
The last column of  Table~\ref{t:sensib} contains either the reference for the origin of the reaction rate
or the standard isotopic mass fraction value obtained with the standard values of the reaction
rates.
In many cases, even if the rate uncertainties have not been explicitly calculated, it is
clear that the error bars are much
lower than the three orders of magnitudes assumed here, essentially because direct or indirect experimental information exists and may constrain the cross section. However, to keep the
procedure simple, we postpone this discussion to the next section. There is indeed
little interest to look for accurate estimates of rate uncertainties for reactions which happen to have no impact on the \sbbn\  in our sensitivity study.

\clearpage

\begin{deluxetable}{cccccccc}
\tiny

\tablecaption{List of reactions included in the sensitivity test with references for the corresponding reaction rate
and relative change in isotopic abundances  when significant. (See text.)}

\startdata
\hline
\hline
 &\multicolumn{6}{|c|}{Reaction} & Ref.        \\
\hline
 $i$ &\multicolumn{6}{|c|}{Enhancement factor  $X_i/X_i^0$ } & $X_i^0$       \\
 \hline
 Factors & 0.001 & 0.01 & 0.1 & 10. & 100. & 1000. & $\times<{\sigma}v>$ \\

 \hline
 &\multicolumn{6}{|c|}{He3(t,np)He4        } &CF88          \\
 \hline
 $^{  7}$Be&1.00$\times10^{  0}$&1.00$\times10^{  0}$&1.00$\times10^{  0}$&                       
 9.97$\times10^{ -1}$&9.72$\times10^{ -1}$&7.96$\times10^{ -1}$&2.61$\times10^{ -9}$\\ 
 $^{  8}$B &1.00$\times10^{  0}$&1.00$\times10^{  0}$&1.00$\times10^{  0}$&                       
 9.97$\times10^{ -1}$&9.72$\times10^{ -1}$&7.96$\times10^{ -1}$&1.99$\times10^{-20}$\\ 
 $^{ 11}$B &9.99$\times10^{ -1}$&1.00$\times10^{  0}$&1.00$\times10^{  0}$&                       
 9.98$\times10^{ -1}$&9.72$\times10^{ -1}$&7.90$\times10^{ -1}$&4.16$\times10^{-15}$\\ 
 $^{ 11}$C &1.00$\times10^{  0}$&1.00$\times10^{  0}$&1.00$\times10^{  0}$&                       
 9.97$\times10^{ -1}$&9.72$\times10^{ -1}$&7.89$\times10^{ -1}$&8.41$\times10^{-18}$\\ 
  \hline
 &\multicolumn{6}{|c|}{He3(t,$\gamma$)Li6         } &FK90          \\
 \hline
 $^{  6}$Li&9.97$\times10^{ -1}$&9.97$\times10^{ -1}$&9.97$\times10^{ -1}$&                       
 1.03$\times10^{  0}$&1.31$\times10^{  0}$&4.11$\times10^{  0}$&5.56$\times10^{-14}$\\ 
 $^{ 10}$B &9.98$\times10^{ -1}$&9.98$\times10^{ -1}$&9.98$\times10^{ -1}$&                       
 1.02$\times10^{  0}$&1.22$\times10^{  0}$&3.20$\times10^{  0}$&2.16$\times10^{-20}$\\ 
\hline
 &\multicolumn{6}{|c|}{He4(d,$\gamma$)Li6          } &Ham10         \\
 \hline
 $^{  6}$Li&4.12$\times10^{ -3}$&1.31$\times10^{ -2}$&1.03$\times10^{ -1}$&                       
 9.97$\times10^{  0}$&9.97$\times10^{  1}$&9.95$\times10^{  2}$&5.56$\times10^{-14}$\\ 
 $^{ 10}$B &4.43$\times10^{ -1}$&4.48$\times10^{ -1}$&4.98$\times10^{ -1}$&                       
 6.02$\times10^{  0}$&5.62$\times10^{  1}$&5.57$\times10^{  2}$&2.16$\times10^{-20}$\\

 \hline
 &\multicolumn{6}{|c|}{Li6($\alpha$,$\gamma$)B10         } &CF88          \\
 \hline
 $^{ 10}$B &4.41$\times10^{ -1}$&4.46$\times10^{ -1}$&4.96$\times10^{ -1}$&                       
 6.04$\times10^{  0}$&5.64$\times10^{  1}$&5.60$\times10^{  2}$&2.16$\times10^{-20}$\\

\hline
 &\multicolumn{6}{|c|}{Li7(n,$\gamma$)Li8         } &MF89Hei98     \\
 \hline
 $^{  8}$Li&1.03$\times10^{ -3}$&1.00$\times10^{ -2}$&1.00$\times10^{ -1}$&                       
 1.00$\times10^{  1}$&1.00$\times10^{  2}$&1.00$\times10^{  3}$&4.24$\times10^{-23}$\\ 
 \hline
 &\multicolumn{6}{|c|}{Li7(d,$\gamma$)Be9         } &TALYS$\dag$        \\
 \hline
 $^{  9}$Be&8.31$\times10^{ -1}$&8.33$\times10^{ -1}$&8.48$\times10^{ -1}$&                       
 2.52$\times10^{  0}$&1.77$\times10^{  1}$&1.70$\times10^{  2}$&2.90$\times10^{-18}$\\ 
 $^{ 10}$Be&1.00$\times10^{  0}$&1.00$\times10^{  0}$&1.00$\times10^{  0}$&                       
 1.00$\times10^{  0}$&1.04$\times10^{  0}$&1.42$\times10^{  0}$&7.32$\times10^{-23}$\\ 
 $^{ 10}$B &9.90$\times10^{ -1}$&9.90$\times10^{ -1}$&9.91$\times10^{ -1}$&                       
 1.09$\times10^{  0}$&2.03$\times10^{  0}$&1.14$\times10^{  1}$&2.16$\times10^{-20}$\\ 
 CNO&9.99$\times10^{ -1}$&9.99$\times10^{ -1}$&9.99$\times10^{ -1}$&                       
 1.01$\times10^{  0}$&1.11$\times10^{  0}$&2.10$\times10^{  0}$&4.98$\times10^{-15}$\\ 
\hline
 &\multicolumn{6}{|c|}{Li7(d,n)2He4        } &Boyd93$\dag$         \\
 \hline
 $^{ 10}$Be&1.00$\times10^{  0}$&1.00$\times10^{  0}$&1.00$\times10^{  0}$&                       
 9.98$\times10^{ -1}$&9.72$\times10^{ -1}$&7.93$\times10^{ -1}$&7.32$\times10^{-23}$\\ 
 CNO&1.66$\times10^{  0}$&1.65$\times10^{  0}$&1.55$\times10^{  0}$&                       
 2.80$\times10^{ -1}$&5.99$\times10^{ -2}$&2.18$\times10^{ -2}$&4.98$\times10^{-15}$\\ 
 \hline
 &\multicolumn{6}{|c|}{Li7(t,$\gamma$)Be10        } &TALYS        \\
 \hline
 $^{ 10}$Be&3.89$\times10^{ -3}$&1.29$\times10^{ -2}$&1.03$\times10^{ -1}$&                       
 9.97$\times10^{  0}$&9.97$\times10^{  1}$&9.97$\times10^{  2}$&7.32$\times10^{-23}$\\ 
 \hline
 &\multicolumn{6}{|c|}{Li7(t,n)Be9         } &Bru90$\dag$         \\
 \hline
 $^{  9}$Be&5.24$\times10^{ -1}$&5.28$\times10^{ -1}$&5.71$\times10^{ -1}$&                       
 5.29$\times10^{  0}$&4.82$\times10^{  1}$&4.77$\times10^{  2}$&2.90$\times10^{-18}$\\ 
 $^{ 10}$Be&9.99$\times10^{ -1}$&9.99$\times10^{ -1}$&9.99$\times10^{ -1}$&                       
 1.01$\times10^{  0}$&1.12$\times10^{  0}$&2.23$\times10^{  0}$&7.32$\times10^{-23}$\\ 
 $^{ 10}$B &9.67$\times10^{ -1}$&9.67$\times10^{ -1}$&9.70$\times10^{ -1}$&                       
 1.30$\times10^{  0}$&4.29$\times10^{  0}$&3.42$\times10^{  1}$&2.16$\times10^{-20}$\\ 
 CNO&9.88$\times10^{ -1}$&9.89$\times10^{ -1}$&9.90$\times10^{ -1}$&                       
 1.10$\times10^{  0}$&2.14$\times10^{  0}$&1.17$\times10^{  1}$&4.98$\times10^{-15}$\\ 
\hline
 &\multicolumn{6}{|c|}{Li7(t,2n)2He4       } &CF88\&MF89     \\
 \hline
 CNO&1.00$\times10^{  0}$&1.00$\times10^{  0}$&1.00$\times10^{  0}$&                       
 9.91$\times10^{ -1}$&9.13$\times10^{ -1}$&5.32$\times10^{ -1}$&4.98$\times10^{-15}$\\ 
 \hline
 &\multicolumn{6}{|c|}{Li7(He3,$\gamma$)B10       } &TALYS        \\
 \hline
 $^{ 10}$B &9.93$\times10^{ -1}$&9.93$\times10^{ -1}$&9.94$\times10^{ -1}$&                       
 1.06$\times10^{  0}$&1.70$\times10^{  0}$&8.08$\times10^{  0}$&2.16$\times10^{-20}$\\ 
 \hline
 &\multicolumn{6}{|c|}{Li7(He3,p)Be9       } &TALYS        \\
 \hline
 $^{  9}$Be&9.96$\times10^{ -1}$&9.96$\times10^{ -1}$&9.96$\times10^{ -1}$&                       
 1.04$\times10^{  0}$&1.45$\times10^{  0}$&5.49$\times10^{  0}$&2.90$\times10^{-18}$\\ 
 $^{ 10}$B &9.98$\times10^{ -1}$&9.98$\times10^{ -1}$&9.98$\times10^{ -1}$&                       
 1.02$\times10^{  0}$&1.19$\times10^{  0}$&2.92$\times10^{  0}$&2.16$\times10^{-20}$\\

\hline
 \tablebreak
\hline

 &\multicolumn{6}{|c|}{Li8(n,$\gamma$)Li9         } &Rau94         \\
 \hline
 CNO&9.99$\times10^{ -1}$&9.99$\times10^{ -1}$&9.99$\times10^{ -1}$&                       
 1.01$\times10^{  0}$&1.06$\times10^{  0}$&1.62$\times10^{  0}$&4.98$\times10^{-15}$\\ 
 \hline
 &\multicolumn{6}{|c|}{Li8(t,n)Be10        } &TALYS        \\
 \hline
 CNO&1.00$\times10^{  0}$&1.00$\times10^{  0}$&1.00$\times10^{  0}$&                       
 1.00$\times10^{  0}$&1.02$\times10^{  0}$&1.23$\times10^{  0}$&4.98$\times10^{-15}$\\ 
 \hline

 &\multicolumn{6}{|c|}{Li8($\alpha$,$\gamma$)B12         } &TALYS        \\
 \hline
 CNO&9.99$\times10^{ -1}$&9.99$\times10^{ -1}$&9.99$\times10^{ -1}$&                       
 1.01$\times10^{  0}$&1.11$\times10^{  0}$&2.15$\times10^{  0}$&4.98$\times10^{-15}$\\ 
 \hline
 &\multicolumn{6}{|c|}{Li8($\alpha$,n)B11         } &Miz01$\dag$          \\
 \hline
 CNO&8.92$\times10^{ -1}$&8.93$\times10^{ -1}$&9.03$\times10^{ -1}$&                       
 1.97$\times10^{  0}$&1.12$\times10^{  1}$&7.81$\times10^{  1}$&4.98$\times10^{-15}$\\ 
 \hline

 &\multicolumn{6}{|c|}{Li9($\alpha$,n)B12         } &TALYS        \\
 \hline
 CNO&9.99$\times10^{ -1}$&9.99$\times10^{ -1}$&9.99$\times10^{ -1}$&                       
 1.01$\times10^{  0}$&1.08$\times10^{  0}$&1.73$\times10^{  0}$&4.98$\times10^{-15}$\\ 
 \hline

 &\multicolumn{6}{|c|}{Be7(d,p)2He4        } &CF88          \\
 \hline
 $^{  7}$Li&1.01$\times10^{  0}$&1.01$\times10^{  0}$&1.00$\times10^{  0}$&                       
 9.57$\times10^{ -1}$&7.36$\times10^{ -1}$&5.05$\times10^{ -1}$&1.54$\times10^{-10}$\\ 
 $^{  8}$Li&1.01$\times10^{  0}$&1.01$\times10^{  0}$&1.00$\times10^{  0}$&                       
 9.57$\times10^{ -1}$&7.36$\times10^{ -1}$&5.05$\times10^{ -1}$&4.24$\times10^{-23}$\\ 
 $^{  7}$Be&1.01$\times10^{  0}$&1.01$\times10^{  0}$&1.01$\times10^{  0}$&                       
 9.23$\times10^{ -1}$&5.27$\times10^{ -1}$&1.13$\times10^{ -1}$&2.61$\times10^{ -9}$\\ 
 $^{  9}$Be&1.01$\times10^{  0}$&1.01$\times10^{  0}$&1.01$\times10^{  0}$&                       
 9.46$\times10^{ -1}$&6.67$\times10^{ -1}$&3.77$\times10^{ -1}$&2.90$\times10^{-18}$\\ 
 $^{ 10}$Be&1.01$\times10^{  0}$&1.00$\times10^{  0}$&1.00$\times10^{  0}$&                       
 9.58$\times10^{ -1}$&7.43$\times10^{ -1}$&5.19$\times10^{ -1}$&7.32$\times10^{-23}$\\ 
 $^{  8}$B &1.01$\times10^{  0}$&1.01$\times10^{  0}$&1.01$\times10^{  0}$&                       
 9.23$\times10^{ -1}$&5.27$\times10^{ -1}$&1.14$\times10^{ -1}$&1.99$\times10^{-20}$\\ 
 $^{ 10}$B &1.00$\times10^{  0}$&1.00$\times10^{  0}$&1.00$\times10^{  0}$&                       
 9.68$\times10^{ -1}$&8.05$\times10^{ -1}$&6.34$\times10^{ -1}$&2.16$\times10^{-20}$\\ 
 $^{ 11}$B &1.01$\times10^{  0}$&1.01$\times10^{  0}$&1.01$\times10^{  0}$&                       
 9.32$\times10^{ -1}$&5.49$\times10^{ -1}$&1.06$\times10^{ -1}$&4.16$\times10^{-15}$\\ 
 $^{ 11}$C &1.01$\times10^{  0}$&1.01$\times10^{  0}$&1.01$\times10^{  0}$&                       
 9.32$\times10^{ -1}$&5.49$\times10^{ -1}$&1.06$\times10^{ -1}$&8.41$\times10^{-18}$\\ 
  \hline
 &\multicolumn{6}{|c|}{Be7(t,$\gamma$)B10         } &TALYS        \\
 \hline
 $^{ 10}$B &6.81$\times10^{ -1}$&6.84$\times10^{ -1}$&7.13$\times10^{ -1}$&                       
 3.87$\times10^{  0}$&3.26$\times10^{  1}$&3.20$\times10^{  2}$&2.16$\times10^{-20}$\\ 
 \hline
 &\multicolumn{6}{|c|}{Be7(t,p)Be9         } &TALYS $\dag$        \\
 \hline
 $^{  9}$Be&6.51$\times10^{ -1}$&6.54$\times10^{ -1}$&6.85$\times10^{ -1}$&                       
 4.15$\times10^{  0}$&3.56$\times10^{  1}$&3.45$\times10^{  2}$&2.90$\times10^{-18}$\\ 
 $^{ 10}$Be&9.99$\times10^{ -1}$&9.99$\times10^{ -1}$&9.99$\times10^{ -1}$&                       
 1.01$\times10^{  0}$&1.11$\times10^{  0}$&2.14$\times10^{  0}$&7.32$\times10^{-23}$\\ 
 $^{ 10}$B &9.32$\times10^{ -1}$&9.33$\times10^{ -1}$&9.39$\times10^{ -1}$&                       
 1.61$\times10^{  0}$&7.72$\times10^{  0}$&6.79$\times10^{  1}$&2.16$\times10^{-20}$\\

\hline
 &\multicolumn{6}{|c|}{Be9(d,n)B10         } &TALYS        \\
 \hline
 $^{ 10}$B &9.98$\times10^{ -1}$&9.98$\times10^{ -1}$&9.98$\times10^{ -1}$&                       
 1.02$\times10^{  0}$&1.23$\times10^{  0}$&3.33$\times10^{  0}$&2.16$\times10^{-20}$\\ 
 \hline
 &\multicolumn{6}{|c|}{Be9(d,p)Be10        } &TALYS        \\
 \hline
 $^{ 10}$Be&9.97$\times10^{ -1}$&9.97$\times10^{ -1}$&9.97$\times10^{ -1}$&                       
 1.03$\times10^{  0}$&1.28$\times10^{  0}$&3.78$\times10^{  0}$&7.32$\times10^{-23}$\\

  \hline
 &\multicolumn{6}{|c|}{Be10(p,$\alpha$)Li7        } &TALYS        \\
 \hline
 $^{ 10}$Be&6.18$\times10^{  2}$&1.53$\times10^{  2}$&1.23$\times10^{  1}$&                       
 8.70$\times10^{ -2}$&8.23$\times10^{ -3}$&8.13$\times10^{ -4}$&7.32$\times10^{-23}$\\ 
 \hline
 &\multicolumn{6}{|c|}{Be10($\alpha$,n)C13        } &TALYS        \\
 \hline
 CNO&1.00$\times10^{  0}$&1.00$\times10^{  0}$&1.00$\times10^{  0}$&                       
 1.00$\times10^{  0}$&1.03$\times10^{  0}$&1.28$\times10^{  0}$&4.98$\times10^{-15}$\\

  \hline
 &\multicolumn{6}{|c|}{B11(n,$\gamma$)B12         } &Rau94$\dag$          \\
 \hline
 CNO&9.10$\times10^{ -1}$&9.11$\times10^{ -1}$&9.19$\times10^{ -1}$&                       
 1.81$\times10^{  0}$&9.91$\times10^{  0}$&8.77$\times10^{  1}$&4.98$\times10^{-15}$\\ 
 \hline
 
 &\multicolumn{6}{|c|}{B11(d,n)C12         } &TALYS $\dag$        \\
 \hline
 CNO&7.04$\times10^{ -1}$&7.06$\times10^{ -1}$&7.33$\times10^{ -1}$&                       
 3.67$\times10^{  0}$&3.02$\times10^{  1}$&2.80$\times10^{  2}$&4.98$\times10^{-15}$\\ 
 \hline
 &\multicolumn{6}{|c|}{B11(d,p)B12         } &TALYS  $\dag$       \\
 \hline
 CNO&9.92$\times10^{ -1}$&9.92$\times10^{ -1}$&9.92$\times10^{ -1}$&                       
 1.08$\times10^{  0}$&1.83$\times10^{  0}$&9.33$\times10^{  0}$&4.98$\times10^{-15}$\\ 
 \hline
 &\multicolumn{6}{|c|}{B11(t,n)C13         } &TALYS        \\
 \hline
 CNO&9.99$\times10^{ -1}$&9.99$\times10^{ -1}$&9.99$\times10^{ -1}$&                       
 1.01$\times10^{  0}$&1.12$\times10^{  0}$&2.17$\times10^{  0}$&4.98$\times10^{-15}$\\

 \hline
 \tablebreak
 \hline

 &\multicolumn{6}{|c|}{C11(n,$\gamma$)C12         } &Rau94         \\
 \hline
 CNO&9.99$\times10^{ -1}$&9.99$\times10^{ -1}$&9.99$\times10^{ -1}$&                       
 1.01$\times10^{  0}$&1.08$\times10^{  0}$&1.75$\times10^{  0}$&4.98$\times10^{-15}$\\ 
 \hline
 &\multicolumn{6}{|c|}{C11(n,$\alpha$)2He4        } &Rau94$\dag$          \\
 \hline
 $^{ 11}$B &1.16$\times10^{  0}$&1.16$\times10^{  0}$&1.15$\times10^{  0}$&                       
 4.02$\times10^{ -1}$&1.16$\times10^{ -2}$&1.63$\times10^{ -4}$&4.16$\times10^{-15}$\\ 
 $^{ 11}$C &1.16$\times10^{  0}$&1.16$\times10^{  0}$&1.15$\times10^{  0}$&                       
 4.01$\times10^{ -1}$&1.11$\times10^{ -2}$&2.77$\times10^{ -6}$&8.41$\times10^{-18}$\\ 
   \hline

 &\multicolumn{6}{|c|}{C11(d,p)C12         } &TALYS$\dag$         \\
 \hline
 CNO&9.94$\times10^{ -1}$&9.94$\times10^{ -1}$&9.95$\times10^{ -1}$&                       
 1.05$\times10^{  0}$&1.55$\times10^{  0}$&5.67$\times10^{  0}$&4.98$\times10^{-15}$\\

 \hline
 &\multicolumn{6}{|c|}{C12(t,$\alpha$)B11         } &TALYS        \\
 \hline
 CNO&1.00$\times10^{  0}$&1.00$\times10^{  0}$&1.00$\times10^{  0}$&                       
 9.97$\times10^{ -1}$&9.68$\times10^{ -1}$&7.49$\times10^{ -1}$&4.98$\times10^{-15}$\\ 
 \hline

 &\multicolumn{6}{|c|}{C13(d,$\alpha$)B11         } &TALYS        \\
 \hline
 CNO&1.00$\times10^{  0}$&1.00$\times10^{  0}$&1.00$\times10^{  0}$&                       
 9.63$\times10^{ -1}$&8.42$\times10^{ -1}$&7.52$\times10^{ -1}$&4.98$\times10^{-15}$\\ 
 \hline
 
\enddata
\\
$\dag$ Reaction rate re-evaluated in Section~\ref{s:improved}.
\label{t:sensib}
\end{deluxetable}

\clearpage

The examination of Table~\ref{t:sensib} shows that only a few reactions have a
strong impact on the CNO or LiBeB productions. We did not consider reactions whose 
only impact would be on \dix\ (or $^{10}$Be), because its abundance would anyway 
remains negligible when compared to \onz.

Note that even with a factor 10$^3$ rate increase we have found no
$^7$Li or $^7$Be $n$-, $p$-, $d$-, \tro- or $\alpha$-induced reactions that   would
significantly reduce the $^7$Li+$^7$Be abundance as suggested
by \citet{Cha11}, except for the $^7$Be(d,p) reaction already considered by
\citet{Coc04,Ang05,Cyb09}.

Reactions affecting the \six\ nucleosynthesis 
are $^4$He(d,$\gamma)^6$Li and to a much lower extent $^3$He(t,$\gamma)^6$Li.   
The former has recently been experimentally re-investigated \citep{Ham10} and its
rate uncertainty should not exceed some 40\%. The rate of the latter has been
calculated by \citet{FK90} without providing an estimate of the associated uncertainty
that should, in any case, be much lower that the factor  of 1000 needed.

The \neu\ nucleosynthesis is sensitive to the $^7$Li(t,n)$^9$Be 
reaction \citep{Boy89,Bru91,Bar91} but also to the $^7$Li(d,$\gamma)^9$Be,  
$^7$Be(t,p)$^9$Be reactions and to a lower extent to the $^7$Li($^3$He,p)$^9$Be.  
These rates are discussed in Section~\ref{s:improved}.

The \onz\ production could be drastically reduced if the $^{11}$C(n,$\alpha$)2$^4$He 
reaction rate (Section~\ref{s:improved}) was higher. 

\begin{figure}[h]
\plotone{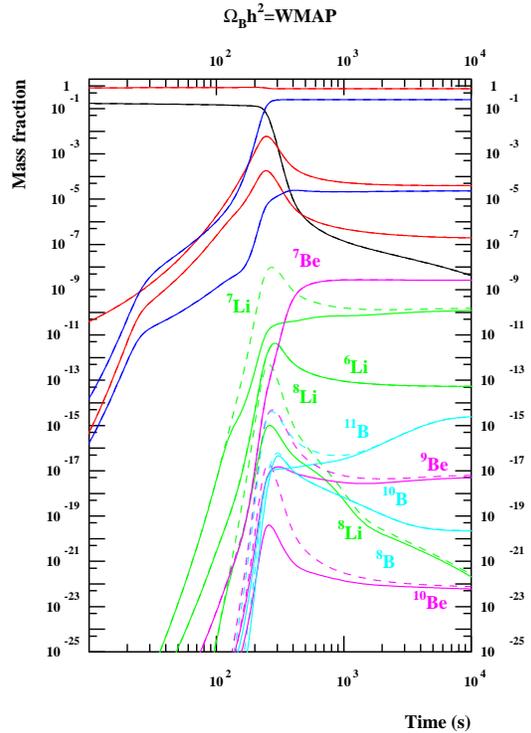}
\caption{(Color online) \sbbn\ production of H, He, Li, Be and B isotopes with the $^7$Li(d,n)2$^4$He reaction rate
from \citet{Boy93} (dashed lines, corresponding to  Fig.~\ref{f:cnotime1}) and with the same rate multiplied by a factor
of 1000 (solid lines).}
\label{f:cnotime3}
\end{figure}

The CNO production is significantly sensitive (more than by a factor of about 2) to several reaction rates. 
In particular, these include:
$^7$Li(d,n)2$^4$He, $^7$Li(t,n)$^9$Be,  $^8$Li($\alpha$,n)$^{11}$B,
$^{11}$B(n,$\gamma)^{12}$C, $^{11}$B(d,n)$^{12}$C, 
$^{11}$B(d,p)$^{12}$B as well as $^{11}$C(d,p)$^{12}$C.
We re-evaluate their rates in Section~\ref{s:improved}.
The impact of $^7$Li(d,n)2$^4$He is unexpected and should be compared to the 
influence of $^1$H(n,$\gamma)^2$H on \sep. Indeed Fig.~\ref{f:cnotime3} shows
the effect of increasing the $^7$Li(d,n)2$^4$He reaction rate by a factor of 1000.
Even though the {\em final abundances are left unchanged}, the peak \sep\ abundance
at $t\approx$200~s is reduced by a factor of about 100. A similar evolution is followed
by the $^8$Li and CNO abundances (not shown on the figure).

\begin{figure*}[h]
\epsscale{2.2}\plotone{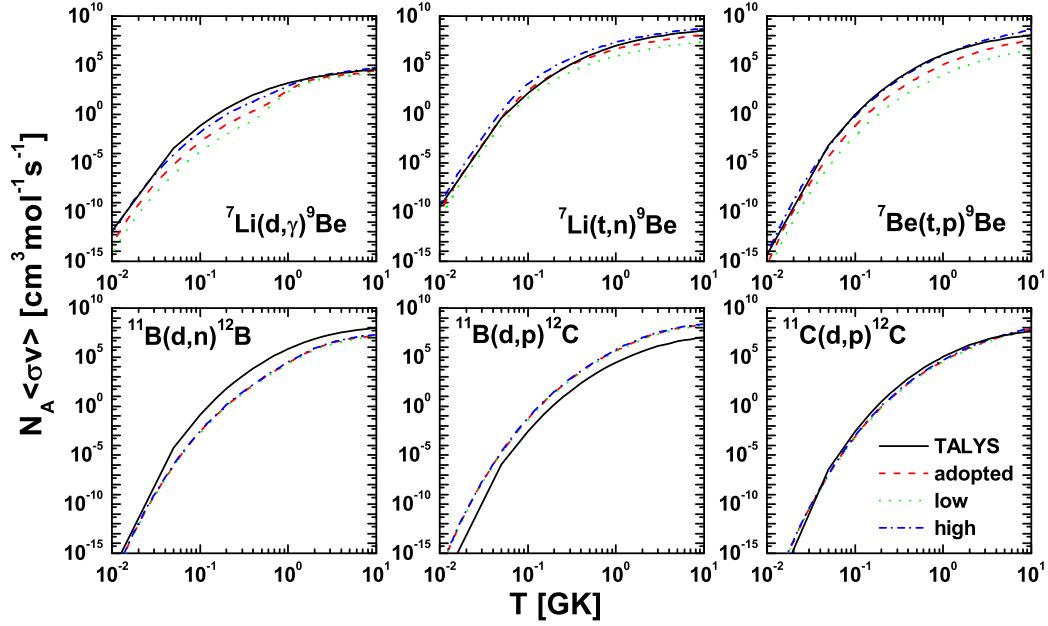}
\caption{(Color online) Comparison of the newly determined rates (dashed line) and their estimated uncertainties (dash-dot and dotted lines) with TALYS predictions (solid line) for the six reactions  $^7$Li(d,$\gamma)^9$Be, $^7$Li(t,n)$^9$B, $^7$Be(t,p)$^9$Be,  $^{11}$B(d,n)$^{12}$C, $^{11}$B(d,p)$^{12}$B, $^{11}$C(d,p)$^{12}$C.}
\label{f:newrates1}
\end{figure*}

\begin{figure*}[h]
\epsscale{1.8}\plotone{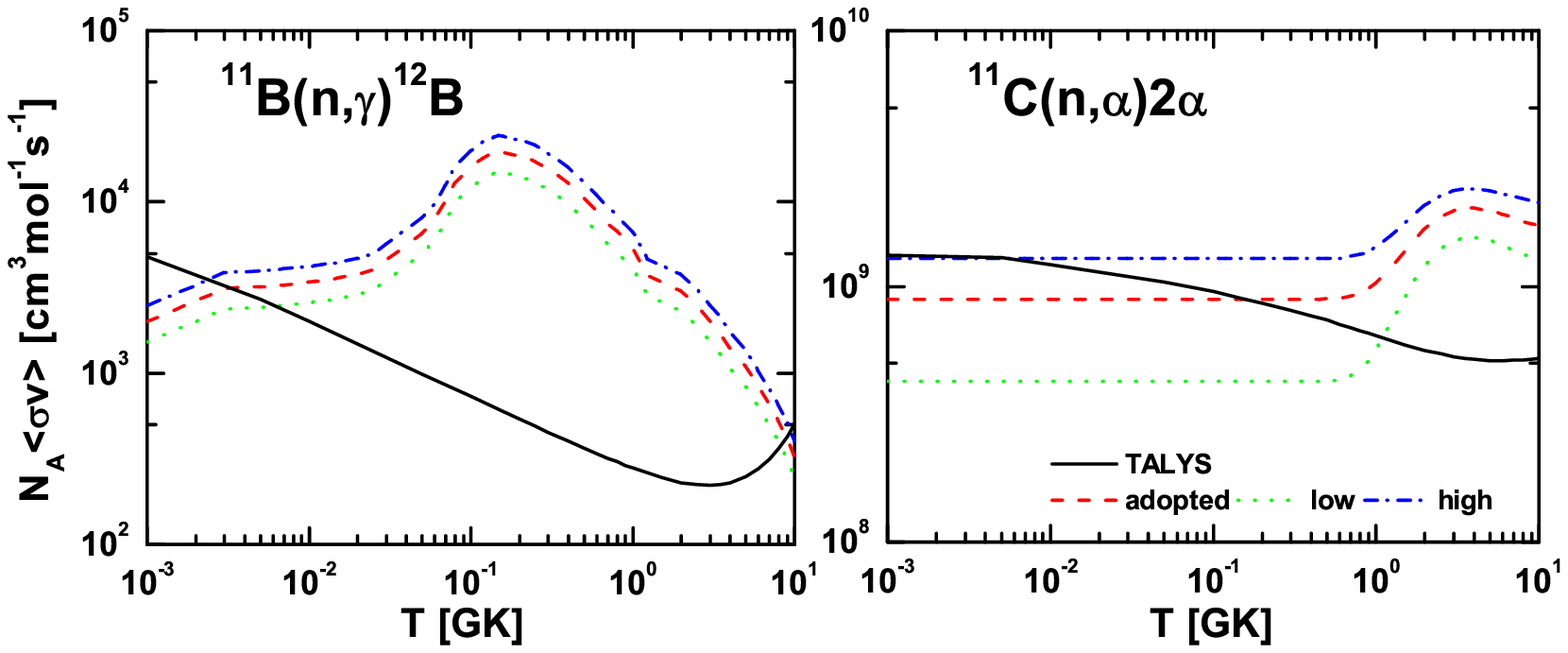}
\caption{(Color online) Same as Fig.~\ref{f:newrates1} for $^{11}$B(n,$\gamma$)$^{12}$C and  $^{11}$C(n,$\alpha$)2$\alpha$.}
\label{f:newrates2}
\end{figure*}

\subsection{Improvement of some critical reaction rates}
\label{s:improved}

In this Section, the above-mentioned critical reactions are analyzed and their rates re-evaluated on the basis of suited reaction models. In addition, realistic uncertainties affecting these rates are estimated in order to provide realistic predictions for the BBN. Since each reaction represents a specific case dominated by a specific reaction mechanism, they are analyzed and evaluated separately below.

\subsubsection{$^7$Li(d,$\gamma)^9$Be affecting \neu} 

The total reaction rate consists  of  two contributions, namely a resonance and a direct part. The direct contribution is obtained by a numerical integration from the experimentally known $S$-factor \citep{li7dg}. The corresponding upper and lower limits are estimated by multiplying the S-factor by a factor 10 and 0.1, respectively. The resonance contribution  is estimated on the basis of Eqs.~(11) and (14) in the NACRE evaluation \citep{NACRE} where the resonance parameters and their uncertainties for the compound system $^9$Be are taken from the RIPL-3 database \citep{ripl3}. The final rate with the estimated uncertainties are shown and compared with TALYS predictions in Fig.~\ref{f:newrates1}.

\subsubsection{$^7$Li(d,n)2~$^4$He affecting CNO} 

\begin{figure}[h*]
\epsscale{1.0}\plotone{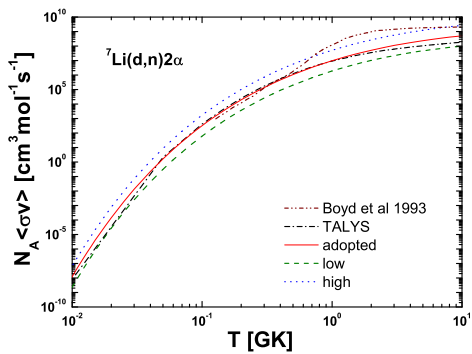}
\caption{(Color online) Estimated rates for $^7$Li(d,n)2$^{4}$He and comparison with TALYS and \citet{Boy93} rates.}
\label{f:newli7dn}
\end{figure}

Both the resonant and direct mechanisms contribute to the total reaction rate. The resonance part is calculated by Eqs. (11) and (14) of ~\citet{NACRE}, where the lowest 4 resonances in the $^7$Li(d,n)2$\alpha$ reaction centre-of-mass system are considered, the corresponding 
resonant parameters being taken from RIPL-3 \cite{ripl3}. For the direct part, the contribution is  obtained by a numerical integration with a constant S-factor of 150 MeV~b is considered for the upper limit \cite{b11dn-4} and of 5.4 MeV~b for the lower limit \cite{Sabo06}. The recommended rate is obtained by the geometrical means of the lower and upper limits of the total rate. 
The final rate with the estimated uncertainties are shown and compared with the TALYS and \citet{Boy93} rates in Fig.~\ref{f:newli7dn}.

\subsubsection{$^7$Li(t,n)$^9$Be affecting CNO and \neu} 

To estimate the $^7$Li(t,n)$^9$B rate, experimental data from \citet{Bru91}, as well as theoretical calculations from \citet{li7tn-2} are considered. 
More precisely, the lower limit of the total reaction rate is obtained from the theoretical analysis of \citet{li7tn-2} based on the experimental determination of the $^7$Li(t,$n_{0}$)$^9$B cross section by \citet{Bru91}. The upper limit is assumed to be a factor of 25 larger than the lower limit. This factor corresponds to the ratio between the $^7$Li(t,$n_{tot}$)$^9$B and $^7$Li(t,$n_{0}$)$^9$B cross sections determined experimentally by  \citet{Bru91} in the 0.20 to 1.4~MeV energy region. The final rates are shown in Fig.~\ref{f:newrates1}.


\subsubsection{$^8$Li($\alpha$,n)$^{11}$B affecting CNO}

For this reaction, various experimental informations exist and may constrain the determination of the reaction rate. In particular, the $^8$Li($\alpha$,$n_{tot}$)$^{11}$B measurements of \citet{li8an-3,li8an-4} above typically 0.6~MeV are used to estimate the upper limit of the cross section, considering a constant $S$-factor at energies below 0.6~MeV. This rate is also used as the recommended rate.

\begin{figure}[h*]
\epsscale{1.0}\plotone{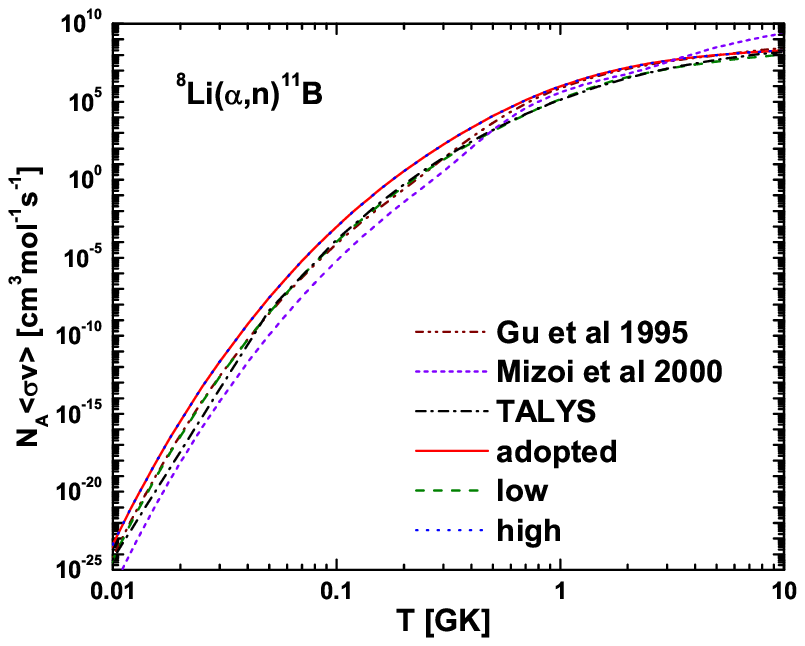}
\caption{(Color online) Adopted rates for $^8$Li($\alpha$,n)$^{11}$B and comparison with TALYS , \citet{Guetal95}  and \citet{Miz00}.}
\label{f:newli8an}
\end{figure}

As far as the lower limit is concerned, experimental constraints are taken from the measurements of \citet{li8an-2} which are significantly lower than those obtained by \citet{li8an-3,li8an-4}. At energies below 0.75~MeV, down to 0.4~MeV, the ground-state experimental cross section  $^8$Li($\alpha$,$n_{0}$)$^{11}$B \citep{li8an-1,li8an-2} provide a lower limit of the cross section, while at energies below 0.4~MeV, a constant $S$-factor is assumed and extrapolated from the \citet{li8an-1} and \citet{li8an-2} data. The final rates are shown in Fig.~\ref{f:newli8an} and compared with  \citet{Guetal95}  and \citet{Miz00}.

\subsubsection{$^7$Be(t,p)$^9$Be affecting \neu} 

The total rate also consists of a resonant plus a direct contribution.  Since no experimental data is available, information on the direct contribution is taken from the mirror $^7$Li($^3$He,p)$^9$Be reaction as given by \citet{li7tn-2}. The upper and lower limits are estimated by multiplying the corresponding S-factor by factors of 10 and 0.1, respectively. The same procedure as used for $^7$Li(d,$\gamma)^9$Be (see above) is followed to determine the resonance component, the resonant parameters of the compound nuclei $^{10}$B as well as their uncertainties being taken from the RIPL-3 library \citep{ripl3}. The final rates are shown in Fig.~\ref{f:newrates1}.

\subsubsection{$^{11}$B(n,$\gamma$)$^{12}$C affecting CNO} 

A recent experimental determination of the $^{11}$B(n,$\gamma$)$^{12}$C cross section was obtained by \citet{Lee10}, including upper and lower limits (see Fig.~\ref{f:newrates2}). Those are considered in the present study. 

\subsubsection{$^{11}$B(d,n)$^{12}$C affecting CNO} 

\begin{figure}[h*]
\epsscale{1.0}\plotone{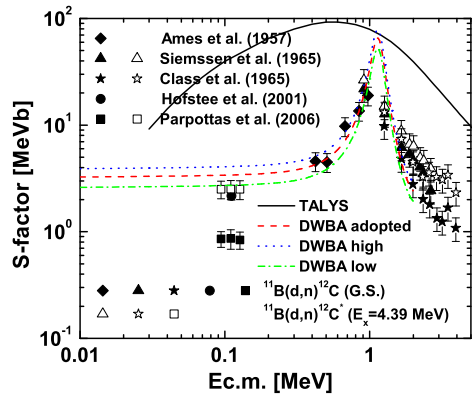}
\caption{(Color online) $^{11}$B(d,n)$^{12}$C experimental and estimated S-factor. The solid line corresponds to the  TALYS prediction.}
\label{f:b11dn}
\end{figure}

For energies $E_{cm}$ ranging between 0.1 and 5 MeV, five sets of experimental data are available \citep{b11dn-1,b11dn-2,b11dn-3,b11dn-4,b11dn-5}. A theoretical Distorted Wave Born Approximation (DWBA) analysis and extrapolation is performed to estimate the S-factors below $E_{c.m.}$=0.1 MeV, as detailed in \citet{NACRE2,NACRE2b}. The DWBA results (along with a 20\% uncertainty) are compared with experimental data in Fig.~\ref{f:b11dn}. The final rates are computed by a numerical integration taking into account the DWBA estimates below $E_{c.m.}$=0.5 MeV and the experimental data with their error bars above (Fig.~\ref{f:b11dn}).

\subsubsection{$^{11}$B(d,p)$^{12}$B affecting CNO} 

\begin{figure}[h*]
\epsscale{1.0}\plotone{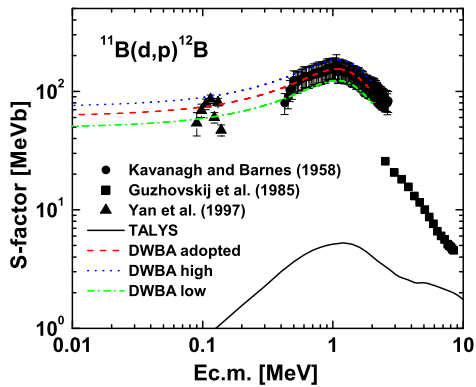}
\caption{(Color online) $^{11}$B(d,p)$^{12}$B experimental and estimated S-factor. The solid line corresponds to the  TALYS prediction.}
\label{f:b11dp}
\end{figure}

Three sets of experimental data have been reported \citep{b11dp-1,b11dp-2,b11dp-3} for energies $E_{cm}$ ranging from  0.1~MeV  to about 10~MeV. To extrapolate the S-factors below $E_{c.m.}$=0.1 MeV, a DWBA evaluation is performed, as described above. The adopted results with an artificial 20$\%$ uncertainties are shown in Fig.~\ref{f:b11dp}  along with the available experimental data. The theoretical DWBA results  below $E_{cm}$ =0.4 MeV  as well as the experimental data with the error bars above $E_{cm}$ =0.4 MeV  are used to estimate the final rates (see Fig.~\ref{f:b11dp}).

\subsubsection{$^{11}$C(n,$\alpha$)2$\alpha$ affecting \onz} 

The total reaction rate consists of two contributions, namely a resonance and a direct component. Concerning the resonant contribution, the calculation is performed on the basis of Eqs. (11) and (14) of the NACRE compilation \citep{NACRE}, where the resonance parameters and their uncertainties in the compound nucleus $^{12}$C are taken from the RIPL-3 library \citep{ripl3}. The direct component (and the corresponding uncertainties) is evaluated based on the thermal cross sections  \citep{Rau94}.  Thanks to the similar properties between the direct $^{11}$C+n and $^{11}$B+p reaction channels, the thermal cross section is obtained separately by performing DWBA calculations for the $^{11}$C(n,$\alpha$)2$\alpha$ reaction where the Woods-Saxon potential is constrained by experimental data on the mirror reaction $^{11}$B(p,$\alpha$)2$\alpha$, as described in the NACRE 2 compilation \citep{NACRE2b}.

\subsubsection{$^{11}$C(d,p)$^{12}$C affecting CNO} 

The total reaction rate consists of the two resonance and direct contributions. As done for previous reactions, for the resonance contribution, the  parameters and their  uncertainties in the compound nucleus $^{13}$N are taken from the RIPL-3 library \citep{ripl3}. The direct component is obtained by a numerical integration with  astrophysical factors $S(0)$  corresponding to 68.43, 82.12 and 54.75 MeVb  for the recommended, upper and lower limits, respectively. These $S$-factors are taken from the direct component in the mirror reaction channel  $^{11}$B(d,p)$^{12}$B, studied in Sect. 3.2.8. 

Finally, the 6 charged-particle-induced reactions and 2 neutron-induced reactions that have been re-evaluated above are compared in Figs.~\ref{f:newrates1}-\ref{f:newrates2} with TALYS predictions. Some deviations are obtained essentially for the cases (e.g. $^8$Li($\alpha$,n)$^{11}$B) where experimental data exist and have been used to constrain the reaction cross sections. For the other cases (e.g.  $^{11}$C(d,p)$^{12}$C), the new determination is also based on theoretical arguments and minor differences are found in fact.


\section{Results}

\subsection{ Standard BBN and comparison with astrophysical constraints}

Table~\ref{t:yields}   displays the comparison between the spectroscopic observations and our
BBN calculated \hli\ abundances  with i) our Monte--Carlo code including the 13 main nuclear 
reactions \citep{CV10} and  our new code with ii) the 13+2 (for \six) nuclear reactions network,  iii) the present 
extended (424) nuclear reaction network.
The small difference between central 
Monte--Carlo values (CV10 column) and the new 15 reaction network results
can be explained by the former use of the  \obh=0.023 value from \citet{Spe07} rather than
the new WMAP 7-years result  \obh=0.02249 \citep{WMAP}. 
The small difference ($\sim$1-2\%) between the 13 and 424 reactions network can be traced back
to the contribution of the additional reactions in the A$<$8 domain. Nevertheless, Table~\ref{t:sensib}
(where the threshold is at a 20\% abundance variation)
shows that none of these reactions can alleviate the lithium problem.  
We hence confirm the robustness of the standard BBN results and refer to a forthcoming paper 
the discussion of the \hli\ aspects.
This is also true for the \six\ abundance compared to \citet{Ham10}.
In particular, no new effective neutron source has been found that may destroy $^7$Be and
reduce the lithium problem.

The concordance  with the spectroscopic observations is in perfect agreement for deuterium.
Considering the large uncertainty associated with \qua\ observations,
the agreement is also fair. The calculated \tro\ value is close to its galactic value
showing that its abundance has not changed significantly during galactic chemical evolution. On the
contrary, as well-known the \sep (CMB+BBN) calculated abundance is significantly higher than the spectroscopic
observations by a factor of about 3.5. Indeed the extended network cannot bring a sufficient neutron source to
modify the primordial value. The origin of this discrepancy between
CMB+BBN and spectroscopic observations remains an open question.
As mentioned before, \citet{Cha11} have proposed  an efficient $^7$Be destruction as a
possible solution to the lithium problem; this destruction requires new resonances, in particular
associated with the $^{10}$C compound nucleus. In our sensitivity study, we have not seen
such an effect that  would require a rate enhancement by resonances larger than
the three orders of magnitude considered here.

The main nuclear path to CNO \citep[see also][]{Ioc07} proceeds from the $^7$Li($\alpha,\gamma)^{11}$B followed
by    $^{11}$B(p,$\gamma$)$^{12}$C, $^{11}$B(d,n)$^{12}$C, $^{11}$B(d,p)$^{12}$B and 
$^{11}$B(n,$\gamma$)$^{12}$B reactions. An other nucleosynthesis paths start with  
$^7$Li(n,$\gamma)^8$Li($\alpha$,n)$^{11}$B. 
The nuclear flow starting by  $^8$B($\alpha,\gamma)^{12}$N is negligible because it is well known \citep{Her96} that
$^8$B production by $^{7}$Be(p,$\gamma$) is hindered at high temperatures by photodisintegration, so that 
its formation is delayed until lower temperatures are reached (see Fig.~\ref{f:cnotime1}).
\neu\ is produced by  $^7$Li(t,n)$^9$Be and  $^7$Be(t,p)$^9$Be while  final \onz\ is produced by the 
late decay of $^{11}$C (see Fig.~\ref{f:cnotime1}).  

\begin{figure*}[h]
\epsscale{2.0}\plotone{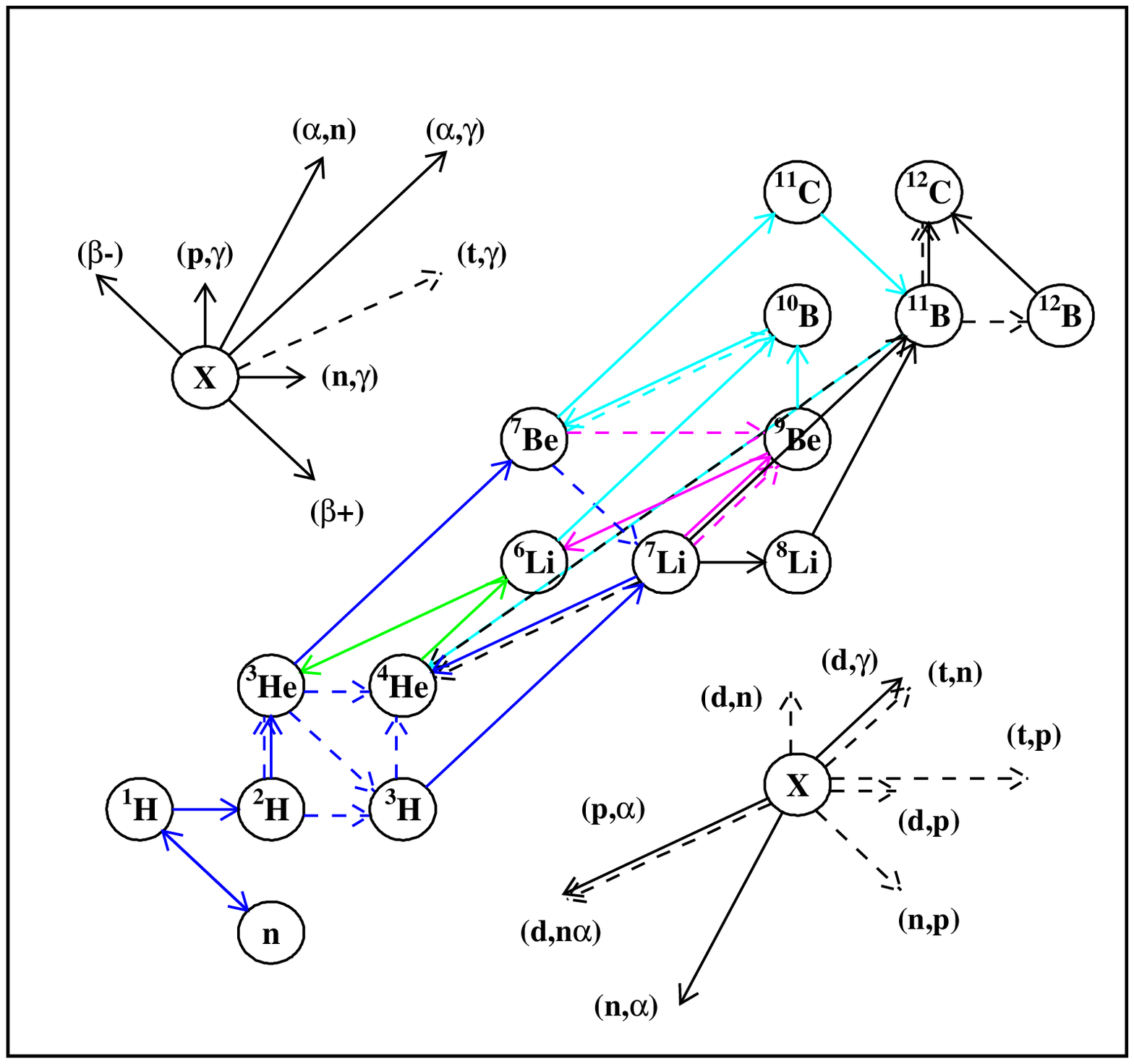}
\caption{ (Color online) Reduced network displaying the important reactions for \hli\ (blue), \six\ (green), 
\neu\ (pink), $^{10,11}$B (cyan) and CNO (black) production. Note that CNO production is via \onz\ but
follows a different path than primordial \onz\ formation through the late time $^{11}$C decay.}
\label{f:reseau}
\end{figure*}

Table~\ref{t:yieldz} compares the BeB  and CNO primordial abundances  calculated with our
extended nuclear network in its initial version with those obtained after its improvement described in 
Section~\ref{s:improved} and with the results of \citet{Ioc07}. As one can see, the
final CNO abundances are very close, i.e CNO/H $\approx$ 5$\times10^{-16}$  though some differences can be found in the isotopic composition. This is particularly visible for $^{14}$C
which is efficiently destroyed in our calculations (see Fig.~\ref{f:cnotime2}) by the 
$^{14}$C(p,$\gamma)^{15}$N and $^{14}$C(d,n)$^{15}$N reactions. $^{14}$C/H
is found to be a factor 10$^4$ more abundant than predicted by  \citet{Ioc07}. The reason for this discrepancy is
unknown as the \citet{Ioc07} network is not given explicitly.
For the sake of comparison, we also performed a nucleosynthesis calculation using \citet{Tho93,Tho94} rates instead of the TALYS rates whenever available
and found a \dix/H abundance higher by a factor of $\sim10^4$ and a CNO abundance higher by a factor of 
$\sim10^2$. We will not discuss these results as they are due to quite  unrealistic
$^{10}$B($\alpha,\gamma)^{14}$N, $^{10}$Be($\alpha,\gamma)^{14}$C and $^{8}$B($\alpha,\gamma)^{12}$N
rates (see~Fig.~\ref{f:tho93beb}).


\begin{table*}[h]
\caption{\label{t:yields} Primordial abundances of H, He and Li isotopes at WMAP7 baryonic density.}
\begin{center}
\begin{tabular}{ccccc}
\hline
  & CV10 & This work    & This work    &Observations \\
 Nb. reactions & 13 (+2) & 15 & 424 & \\
\hline
$Y_p$     &  0.2476$\pm$0.0004       & 0.2475       & 0.2476 & $0.2561\pm 0.0108$  \\
\deu/H   ($ \times10^{-5})$& $2.68\pm0.15$    & 2.64   &     2.59    & 2.82$\pm$0.2 \\
\tro/H    ($ \times10^{-5}$) & 1.05$\pm$0.04    & 1.05      &   1.04    & 1.1 $\pm$0.2        \\
\sep/H ($\times10^{-10}$)  &  5.14$\pm$0.50 &  5.20  & 5.24 & 1.58$\pm$0.31   \\
\six/H ($\times10^{-14}$)& 1.3$^\dag$ & 1.32 & 1.23 & $\sim$1000 (?)\\
\hline
\end{tabular}\\
CV10: \citet{CV10}  using \obh\ from \citet{Spe07}. This work uses the new \citet{WMAP} value.\\
$^\dag$\citet{Ham10} \\
\end{center}
\end{table*}


\begin{table*}[h]
\caption{\label{t:yieldz} Be to CNO primordial abundances by number at WMAP7 baryonic density.}
\begin{center}
\begin{tabular}{ccccc}
\hline
  & IMMPS07& This work &  This work  \\
&  & (Initial*) &  (Improved*)  \\
\hline
\neu/H ($\times10^{-19}$) & 2.5  & 2.24 &      9.60             \\
\dix/H  ($\times10^{-21}$)& & 2.78&   3.00  \\
\onz/H  ($\times10^{-16}$) & 3.9 &  5.86 &  3.05      \\
$^{12}$C/H ($\times10^{-16}$)&4.6&3.56& 5.34\\
$^{13}$C/H ($\times10^{-16}$) &0.90&0.87& 1.41\\
$^{14}$C/H  ($\times10^{-21}$) &13000.&0.96& 1.62\\
$^{14}$N/H ($\times10^{-17}$)&3.7&3.98& 6.76 \\
$^{15}$N/H ($\times10^{-20}$)&&1.32& 2.25 \\
$^{16}$O/H ($\times10^{-20}$) &2.7&5.18& 9.13  \\
CNO/H  ($\times10^{-16}$) &6.00& 4.83&    7.43 \\
\hline
\end{tabular}\\
IMMPS07: \citet{Ioc07} C3 code. \\ 
**'Initial' corresponds to original network before re-evaluations of selected reaction rates 'improved' (see text).
\end{center}
\end{table*}

\section{Conclusion}

We have used an extensive network of more than 400 nuclear reactions whose thermonuclear reaction rates are
adopted from evaluations based on experimental data or were calculated using the TALYS code. It enabled
us to calculate the \sbbn\ production of \six, \neu, \dix, \onz\ and CNO isotopes  more reliably. We performed
a sensitivity study by varying uncertain reaction rates by factors of up to 1000 and down to 0.001. 
In that way, a few reactions that could affect the A$>$7 isotope yields were identified and their rate
 (re-)evaluated using available experimental data and/or theoretical or phenomenological input.
On the basis of these new evaluations the  \neu, \dix, \onz\ and CNO isotope production was found to be 
close to the initial results and in global agreement with previous calculations \citep{Ioc07}.

For most of the few reactions which were identified to have an impact on \sbbn, we were able to
collect sufficient experimental data to derive new reaction rates with associated uncertainties 
much reduced with respect to our initial three orders of magnitude variation. 
(An exception is $^7$Be(t,p)$^9$Be but affecting the  \neu\ production only.)
In some cases these new rates differ from the previous ones by large factors but changes compensating
each other (e.g. $^{11}$B(d,n)$^{12}$C and $^{11}$B(d,p)$^{12}$B) allow us to
confirm the CNO \sbbn\ production of about $0.7\times10^{-14}$ in mass fraction (i.e CNO/H $\approx0.7\times10^{-15}$ ).
Based on the rate uncertainties (a factor of $\lap$10 at BBN temperatures) obtained in Section~\ref{s:improved} and
combining \citep[see][]{ILCCF10a} the corresponding uncertainty factors from Table~\ref{t:sensib} 
we can estimate the uncertainty on CNO production to be of a factor of  $\lap$4. 
This is too small to
have presently an impact on Population III stellar evolution. It is nevertheless a reference value for comparison
with non-\sbbn\ CNO production e.g. in the context of varying constants.

Finally, our extension of the network does not help alleviate the discrepancy between the calculated
and observed  \sep\  abundances. No new late-time neutron source was found that could destroy
\be\ before it decays to \sep. 

Nevertheless we pointed out the unexpected high sensitivity of the CNO 
abundance with respect to the $^7$Li(d,n)2$^4$He reaction rate. A similar situation was found between the \sep\ abundance  and the $^1$H(n,$\gamma)^2$H reaction rate. 
This emphasizes again the complex nature of nucleosynthesis and a posteriori justifies such a sensitivity study:
 the impact of a given reaction being not always predictable, even in the simple case (i.e. homogeneity, no mixing, nor convection,...) of BBN.
 As a consequence, even though very unlikely, the search for a nuclear solution to the lithium problem remains justified.

\acknowledgements
Acknowledgements.
We apologize in advance for unintentionally omitting references to evaluated reaction rates
unknown to us.  S.G. and X.Y. acknowledge the financial support of the "Actions de recherche concert\'ees (ARC)" from the "Communaut\'e fran\c caise de Belgique". S.G is F.N.R.S. research associate.

\appendix

\section{Network}

\begin{deluxetable}{ccc|ccc}

\tablecaption{Network}

\startdata

\hline

Reaction & Ref. & Q (MeV) &Reaction & Ref. & Q (MeV) \\

\hline

$^{1}$n(p,$\gamma$)$^{2}$H  &  And06  &  2.2246  &  $^{1}$H(n,p)$^{2}$H  &  CF88  &  2.2246\\
$^{2}$H(n,$\gamma$)$^{3}$H  &  Nag06  &  6.2572  &  $^{2}$H(p,$\gamma$)$^{3}$He  &  DAACV04  &  5.4935\\
$^{2}$H(d,n)$^{3}$He  &  DAACV04  &  3.2689  &  $^{2}$H(d,p)$^{3}$H  &  DAACV04  &  4.0327\\
$^{2}$H($\alpha,\gamma$)$^{6}$Li  &  Ham10  &  1.4738  &  $^{2}$H(d,$\gamma$)$^{4}$He  &  NACREII  &  23.8465\\
$^{3}$H(t,2n)$^{4}$He  &  CF88  &  11.3321  &  $^{3}$H($\alpha$,n)$^{6}$Li  &  CF88  &  -4.7834\\
$^{3}$H(p,$\gamma$)$^{4}$He  &  Ser04  &  19.8139  &  $^{3}$H(d,n)$^{4}$He  &  DAACV04  &  17.5893\\
$^{3}$H($\alpha,\gamma$)$^{7}$Li  &  DAACV04  &  2.4666  &  $^{3}$He(n,$\gamma$)$^{4}$He  &  ?Wag69  &  20.5776\\
$^{3}$He(t,d)$^{4}$He  &  CF88  &  14.3204  &  $^{3}$He(t,np)$^{4}$He  &  CF88  &  12.0958\\
$^{3}$He(t,$\gamma$)$^{6}$Li  &  FK90  &  15.7942  &  $^{3}$He(n,p)$^{3}$H  &  DAACV04  &  0.7638\\
$^{3}$He(d,p)$^{4}$He  &  DAACV04  &  18.3531  &  $^{3}$He($\alpha,\gamma$)$^{7}$Be  &  Cyb08  &  1.5861\\
$^{3}$He($^3$He,2p)$^{4}$He  &  NACREII  &  12.8596  &  $^{4}$He($\alpha$n,$\gamma$)$^{9}$Be  &  NACRE  &  1.5735\\
$^{4}$He(np,$\gamma$)$^{6}$Li  &  CF88!  &  3.6984  &  $^{4}$He($\alpha\alpha$,$\gamma$$^{12}$C  &  NACRE  &  7.2747\\
$^{4}$He(2n,$\gamma$)$^{6}$He  &  Efr96  &  0.9724  &  $^{6}$Li(n,$\gamma$)$^{7}$Li  &  MF89  &  7.2500\\
$^{6}$Li(d,p)$^{7}$Li  &  MF89  &  5.0254  &  $^{6}$Li(d,n)$^{7}$Be  &  MF89  &  3.3812\\
$^{6}$Li($\alpha,\gamma$)$^{10}$B  &  CF88  &  4.4610  &  $^{6}$Li(p,$\gamma$)$^{7}$Be  &  NACREII  &  5.6057\\
$^{6}$Li(p,$\alpha$)$^{3}$He  &  NACREII  &  4.0196  &  $^{6}$Li($^3$He,p)$^{4}$He  &  TALYS  &  16.8792\\
$^{6}$Li(t,$\gamma$)$^{9}$Be  &  TALYS  &  17.6890  &  $^{6}$Li(t,n)$^{4}$He  &  TALYS  &  16.1155\\
$^{6}$Li(t,p)$^{8}$Li  &  TALYS  &  0.8008  &  $^{6}$Li(t,d)$^{7}$Li  &  TALYS  &  0.9927\\
$^{6}$Li($^3$He,d)$^{7}$Be  &  TALYS  &  0.1123  &  {\bf $^{7}$Li(d,n)2$^{4}$He}  &  {\bf Boy93}  &  15.1227\\
$^{7}$Li(t,2n)$^{4}$He  &  CF88\&  MF89  &  8.8655  &  $^{7}$Li($^3$He,np)$^{4}$He  &  CF88\&  MF89  &  9.6292\\
{\bf $^{7}$Li(t,n)$^{9}$Be}  &  {\bf Bru91}  &  10.4390  &  $^{7}$Li($\alpha$,n)$^{10}$B  &  NACRE  &  -2.7890\\
$^{7}$Li(n,$\gamma$)$^{8}$Li  &  MF89Hei98  &  2.0326  &  $^{7}$Li(d,p)$^{8}$Li  &  MF89  &  -0.1919\\
$^{7}$Li(p,$\alpha$)$^{4}$He  &  DAACV04  &  17.3473  &  $^{7}$Li(p,$\gamma$)$^{4}$He  &  NACREII  &  17.3473\\
{\bf $^{7}$Li(d,$\gamma$)$^{9}$Be}  &  {\bf TALYS}  &  16.6963  &  $^{7}$Li($^3$He,$\gamma$)$^{10}$B  &  TALYS  &  17.7887\\
$^{7}$Li($^3$He,$\alpha$)$^{6}$Li  &  TALYS  &  13.3276  &  $^{7}$Li(t,$\gamma$)$^{10}$Be  &  TALYS  &  17.2512\\
$^{7}$Li($^3$He,p)$^{9}$Be  &  TALYS  &  11.2028  &  $^{7}$Li($^3$He,d)$^{4}$He  &  TALYS  &  11.8538\\
$^{7}$Li($\alpha,\gamma$)$^{11}$B  &  NACREII  &  8.6652  &  $^{8}$Li(d,p)$^{9}$Li  &  Bal95  &  1.8393\\
$^{8}$Li(d,t)$^{7}$Li  &  Has09  &  4.2246  &  $^{8}$Li(n,$\gamma$)$^{9}$Li  &  Rau94  &  4.0639\\
$^{8}$Li(p,n)$^{4}$He  &  Bec92  &  15.3147  &  $^{8}$Li(d,n)$^{9}$Be  &  Bal95  &  14.6636\\
$^{8}$Li(p,$\gamma$)$^{9}$Be  &  TUNL\&  Cam08  &  16.8882  &  $^{8}$Li($\alpha,\gamma$)$^{12}$B  &  TALYS  &  10.0029\\
{\bf $^{8}$Li($\alpha$,n)$^{11}$B}  &  {\bf Miz00}  &  6.6325  &  $^{8}$Li(d,$\gamma$)$^{10}$Be  &  TALYS  &  21.4759\\
$^{8}$Li($^3$He,$\gamma$)$^{11}$B  &  TALYS  &  27.2102  &  $^{8}$Li($^3$He,n)$^{10}$B  &  TALYS  &  15.7560\\
$^{8}$Li($^3$He,p)$^{10}$Be  &  TALYS  &  15.9824  &  $^{8}$Li($^3$He,$\alpha$)$^{7}$Li  &  TALYS  &  18.5450\\
$^{8}$Li(t,$\gamma$)$^{11}$Be  &  TALYS  &  15.7226  &  $^{8}$Li(t,n)$^{10}$Be  &  TALYS  &  15.2186\\
$^{8}$Li($^3$He,d)$^{9}$Be  &  TALYS  &  11.3947  &  $^{8}$Li($^3$He,t)$^{4}$He  &  TALYS  &  16.0784\\
$^{9}$Li(p,t)$^{7}$Li  &  TALYS  &  2.3853  &  $^{9}$Li(d,n)$^{10}$Be  &  TALYS  &  17.4120\\
$^{9}$Li(p,n)$^{9}$Be  &  TALYS  &  12.8244  &  $^{9}$Li($\alpha$,n)$^{12}$B  &  TALYS  &  5.9390\\
$^{9}$Li(p,$\gamma$)$^{10}$Be  &  TALYS  &  19.6366  &  $^{9}$Li($\alpha,\gamma$)$^{13}$B  &  TALYS  &  10.8170\\
$^{9}$Li(d,$\gamma$)$^{11}$Be  &  TALYS  &  17.9160  &  $^{9}$Li($^3$He,$\gamma$)$^{12}$B  &  TALYS  &  26.5166\\
$^{9}$Li($^3$He,n)$^{11}$B  &  TALYS  &  23.1463  &  $^{9}$Li($^3$He,p)$^{11}$Be  &  TALYS  &  12.4225\\
$^{9}$Li($^3$He,$\alpha$)$^{8}$Li  &  TALYS  &  16.5138  &  $^{9}$Li(t,$\gamma$)$^{12}$Be  &  TALYS  &  14.8271\\
$^{9}$Li(t,n)$^{11}$Be  &  TALYS  &  11.6588  &  $^{9}$Li(d,t)$^{8}$Li  &  TALYS  &  2.1934\\
$^{9}$Li($^3$He,d)$^{10}$Be  &  TALYS  &  14.1431  &  $^{9}$Li($^3$He,t)$^{9}$Be  &  TALYS  &  13.5881\\
$^{7}$Be(n,p)$^{7}$Li  &  DAACV04  &  1.6442  &  $^{7}$Be(d,p)$^{4}$He  &  CF88  &  16.7670\\
$^{7}$Be(t,np)$^{4}$He  &  CF88\&  MF89  &  10.5097  &  $^{7}$Be($^3$He,2p)$^{4}$He  &  CF88\&  MF89  &  11.2735\\
$^{7}$Be($^3$He,$\gamma$)$^{10}$C  &  TALYS  &  15.0025  &  $^{7}$Be(n,$\gamma$)$^{4}$He  &  TALYS  &  18.9915\\
$^{7}$Be(t,$\gamma$)$^{10}$B  &  TALYS  &  18.6691  &  {\bf $^{7}$Be(t,p)$^{9}$Be}  &  {\bf TALYS}  &  12.0833\\
$^{7}$Be(t,$\alpha$)$^{6}$Li  &  TALYS  &  14.2081  &  $^{7}$Be(t,d)$^{4}$He  &  TALYS  &  12.7343\\
$^{7}$Be(t,$^3$He)$^{7}$Li  &  TALYS  &  0.8805  &  $^{7}$Be($^3$He,p)2$\alpha$  &  TALYS  &  11.2735\\
$^{7}$Be(p,$\gamma$)$^{8}$B  &  NACREII  &  0.1375  &  $^{7}$Be($\alpha,\gamma$)$^{11}$C  &  NACREII  &  7.5446\\
$^{9}$Be(n,$\gamma$)$^{10}$Be  &  Rau94  &  6.8122  &  $^{9}$Be(p,pn)$^{4}$He  &  NACRE  &  -1.5735\\
$^{9}$Be(t,n)$^{11}$B  &  TALYS  &  9.5582  &  $^{9}$Be($\alpha,\gamma$)$^{13}$C  &  TALYS  &  10.6475\\
$^{9}$Be(d,$\gamma$)$^{11}$B  &  TALYS  &  15.8154  &  $^{9}$Be(d,n)$^{10}$B  &  TALYS  &  4.3613\\
$^{9}$Be(d,p)$^{10}$Be  &  TALYS  &  4.5877  &  $^{9}$Be(d,$\alpha$)$^{7}$Li  &  TALYS  &  7.1503\\
$^{9}$Be($^3$He,$\gamma$)$^{12}$C  &  TALYS  &  26.2788  &  $^{9}$Be($^3$He,n)$^{11}$C  &  TALYS  &  7.5572\\
$^{9}$Be($^3$He,p)$^{11}$B  &  TALYS  &  10.3219  &  $^{9}$Be($^3$He,$\alpha$)$^{4}$He  &  TALYS  &  19.0041\\
$^{9}$Be(t,$\gamma$)$^{12}$B  &  TALYS  &  12.9285  &  $^{9}$Be(t,$\alpha$)$^{8}$Li  &  TALYS  &  2.9257\\
$^{9}$Be(d,t)$^{4}$He  &  TALYS  &  4.6837  &  $^{9}$Be(t,d)$^{10}$Be  &  TALYS  &  0.5550\\
$^{9}$Be($^3$He,d)$^{10}$B  &  TALYS  &  1.0924  &  $^{9}$Be(p,$\gamma$)$^{10}$B  &  NACREII  &  6.5859\\
$^{9}$Be(p,d)$^{4}$He  &  NACREII  &  0.6510  &  $^{9}$Be(p,$\alpha$)$^{6}$Li  &  NACREII  &  2.1249\\
$^{9}$Be($\alpha$,n)$^{12}$C  &  NACREII  &  5.7012  &  $^{10}$Be(n,$\gamma$)$^{11}$Be  &  Rau94  &  0.5040\\
$^{10}$Be($\alpha,\gamma$)$^{14}$C  &  TALYS  &  12.0117  &  $^{10}$Be(p,$\gamma$)$^{11}$B  &  TALYS  &  11.2278\\
$^{10}$Be(p,$\alpha$)$^{7}$Li  &  TALYS  &  2.5626  &  $^{10}$Be($\alpha$,n)$^{13}$C  &  TALYS  &  3.8353\\
$^{10}$Be(d,$\gamma$)$^{12}$B  &  TALYS  &  12.3735  &  $^{10}$Be(d,n)$^{11}$B  &  TALYS  &  9.0032\\
$^{10}$Be(d,$\alpha$)$^{8}$Li  &  TALYS  &  2.3707  &  $^{10}$Be($^3$He,$\gamma$)$^{13}$C  &  TALYS  &  24.4129\\
$^{10}$Be($^3$He,n)$^{12}$C  &  TALYS  &  19.4666  &  $^{10}$Be($^3$He,p)$^{12}$B  &  TALYS  &  6.8800\\
$^{10}$Be($^3$He,$\alpha$)$^{9}$Be  &  TALYS  &  13.7654  &  $^{10}$Be(t,$\gamma$)$^{13}$B  &  TALYS  &  10.9943\\
$^{10}$Be(t,n)$^{12}$B  &  TALYS  &  6.1163  &  $^{10}$Be(t,$\alpha$)$^{9}$Li  &  TALYS  &  0.1773\\
$^{10}$Be(p,t)$^{4}$He  &  TALYS  &  0.0960  &  $^{10}$Be($^3$He,d)$^{11}$B  &  TALYS  &  5.7343\\
$^{10}$Be($^3$He,t)$^{10}$B  &  TALYS  &  0.5374  &  $^{11}$Be(n,$\gamma$)$^{12}$Be  &  Rau94  &  3.1683\\
$^{11}$Be(p,$\alpha$)$^{8}$Li  &  TALYS  &  4.0912  &  $^{11}$Be(p,n)$^{11}$B  &  TALYS  &  10.7238\\
$^{11}$Be($\alpha$,n)$^{14}$C  &  TALYS  &  11.5077  &  $^{11}$Be($\alpha,\gamma$)$^{15}$C  &  TALYS  &  12.7258\\
$^{11}$Be(d,$\gamma$)$^{13}$B  &  TALYS  &  16.7475  &  $^{11}$Be(d,n)$^{12}$B  &  TALYS  &  11.8695\\
$^{11}$Be(d,p)$^{12}$Be  &  TALYS  &  0.9438  &  $^{11}$Be(d,$\alpha$)$^{9}$Li  &  TALYS  &  5.9305\\
$^{11}$Be($^3$He,$\gamma$)$^{14}$C  &  TALYS  &  32.0853  &  $^{11}$Be($^3$He,n)$^{13}$C  &  TALYS  &  23.9089\\
$^{11}$Be($^3$He,p)$^{13}$B  &  TALYS  &  11.2540  &  $^{11}$Be($^3$He,$\alpha$)$^{10}$Be  &  TALYS  &  20.0736\\
$^{11}$Be(p,$\gamma$)$^{12}$B  &  TALYS  &  14.0941  &  $^{11}$Be(t,$\gamma$)$^{14}$B  &  TALYS  &  11.4598\\
$^{11}$Be(t,n)$^{13}$B  &  TALYS  &  10.4903  &  $^{11}$Be(p,t)$^{9}$Be  &  TALYS  &  1.1656\\
$^{11}$Be(p,d)$^{10}$Be  &  TALYS  &  1.7205  &  $^{12}$Be($\alpha,\gamma$)$^{16}$C  &  TALYS  &  13.8079\\
$^{12}$Be($\alpha$,n)$^{15}$C  &  TALYS  &  9.5575  &  $^{12}$Be(d,$\gamma$)$^{14}$B  &  TALYS  &  14.5487\\
$^{12}$Be(d,n)$^{13}$B  &  TALYS  &  13.5792  &  $^{12}$Be($^3$He,$\gamma$)$^{15}$C  &  TALYS  &  30.1351\\
$^{12}$Be($^3$He,n)$^{14}$C  &  TALYS  &  28.9170  &  $^{12}$Be($^3$He,p)$^{14}$B  &  TALYS  &  9.0552\\
$^{12}$Be($^3$He,$\alpha$)$^{11}$Be  &  TALYS  &  17.4093  &  $^{12}$Be(p,$\gamma$)$^{13}$B  &  TALYS  &  15.8038\\
$^{12}$Be(p,n)$^{12}$B  &  TALYS  &  10.9258  &  $^{12}$Be(t,$\gamma$)$^{15}$B  &  TALYS  &  11.0548\\
$^{12}$Be(t,n)$^{14}$B  &  TALYS  &  8.2915  &  $^{12}$Be(p,$\alpha$)$^{9}$Li  &  TALYS  &  4.9868\\
$^{12}$Be(p,t)$^{10}$Be  &  TALYS  &  4.8095  &  $^{8}$B(p,$\gamma$)$^{9}$C  &  Des99Bea01  &  1.3000\\
$^{8}$B($\alpha,\gamma$)$^{12}$N  &  TALYS  &  8.0083  &  $^{8}$B(n,p)$^{4}$He  &  TALYS  &  18.8540\\
$^{8}$B($\alpha$,p)$^{11}$C  &  TALYS  &  7.4071  &  $^{8}$B(d,$\gamma$)$^{10}$C  &  TALYS  &  20.3585\\
$^{8}$B($^3$He,p)$^{10}$C  &  TALYS  &  14.8650  &  $^{8}$B(t,$\gamma$)$^{11}$C  &  TALYS  &  27.2210\\
$^{8}$B(t,n)$^{10}$C  &  TALYS  &  14.1013  &  $^{8}$B(t,p)$^{10}$B  &  TALYS  &  18.5316\\
$^{8}$B(t,$\alpha$)$^{7}$Be  &  TALYS  &  19.6764  &  $^{8}$B(n,$^3$He)$^{6}$Li  &  TALYS  &  1.9748\\
$^{8}$B(n,d)$^{7}$Be  &  TALYS  &  2.0871  &  $^{8}$B(d,$^3$He)$^{7}$Be  &  TALYS  &  5.3560\\
$^{8}$B(t,$^3$He)$^{4}$He  &  TALYS  &  18.0903  &  $^{10}$B($\alpha$,n)$^{13}$N  &  CF88  &  1.0588\\
$^{10}$B($\alpha,\gamma$)$^{14}$N  &  TALYS  &  11.6122  &  $^{10}$B(n,$\gamma$)$^{11}$B  &  TALYS  &  11.4541\\
$^{10}$B($\alpha$,p)$^{13}$C  &  TALYS  &  4.0616  &  $^{10}$B(d,$\gamma$)$^{12}$C  &  TALYS  &  25.1864\\
$^{10}$B(d,n)$^{11}$C  &  TALYs  &  6.4648  &  $^{10}$B(d,p)$^{11}$B  &  TALYS  &  9.2296\\
$^{10}$B(d,$\alpha$)$^{4}$He  &  TALYS  &  17.9117  &  $^{10}$B($^3$He,$\gamma$)$^{13}$N  &  TALYS  &  21.6364\\
$^{10}$B($^3$He,n)$^{12}$N  &  TALYS  &  1.5725  &  $^{10}$B($^3$He,p)$^{12}$C  &  TALYS  &  19.6929\\
$^{10}$B(n,p)$^{10}$Be  &  TALYs  &  0.2263  &  $^{10}$B(t,$\gamma$)$^{13}$C  &  TALYS  &  23.8755\\
$^{10}$B(t,n)$^{12}$C  &  TALYS  &  18.9292  &  $^{10}$B(t,p)$^{12}$B  &  TALYS  &  6.3426\\
$^{10}$B(t,$\alpha$)$^{9}$Be  &  TALYS  &  13.2280  &  $^{10}$B(n,t)$^{4}$He  &  TALYS  &  0.3224\\
$^{10}$B($\alpha$,d)$^{12}$C  &  TALYS  &  1.3399  &  $^{10}$B(p,$^3$He)$^{4}$He  &  TALYS  &  -0.4414\\
$^{10}$B(t,d)$^{11}$B  &  TALYS  &  5.1969  &  $^{10}$B($^3$He,d)$^{11}$C  &  TALYS  &  3.1959\\
$^{10}$B(p,$\gamma$)$^{11}$C  &  NACREII  &  8.6894  &  $^{10}$B(p,$\alpha$)$^{7}$Be  &  NACREII  &  1.1447\\
$^{11}$B(p,n)$^{11}$C  &  NACRE  &  -2.7647  &  {\bf $^{11}$B(n,$\gamma$)$^{12}$B}  &  {\bf Rau94}  &  3.3703\\
$^{11}$B($\alpha$,p)$^{14}$C  &  Wan91  &  0.7840  &  $^{11}$B($\alpha,\gamma$)$^{15}$N  &  Wan91  &  10.9914\\
$^{11}$B(d,$\gamma$)$^{13}$C  &  TALYS  &  18.6786  &  {\bf $^{11}$B(d,n)$^{12}$C}  &  {\bf TALYS}  &  13.7323\\
{\bf $^{11}$B(d,p)$^{12}$B}  &  {\bf TALYS}  &  1.1458  &  $^{11}$B(d,$\alpha$)$^{9}$Be  &  TALYS  &  8.0311\\
$^{11}$B($^3$He,$\gamma$)$^{14}$N  &  TALYS  &  20.7357  &  $^{11}$B($^3$He,n)$^{13}$N  &  TALYS  &  10.1823\\
$^{11}$B($^3$He,p)$^{13}$C  &  TALYS  &  13.1851  &  $^{11}$B($^3$He,$\alpha$)$^{10}$B  &  TALYS  &  9.1235\\
$^{11}$B(t,$\gamma$)$^{14}$C  &  TALYS  &  20.5978  &  $^{11}$B(t,n)$^{13}$C  &  TALYS  &  12.4214\\
$^{11}$B(t,$\alpha$)$^{10}$Be  &  TALYS  &  8.5861  &  $^{11}$B(t,p)$^{13}$B  &  TALYS  &  -0.2335\\
$^{11}$B($^3$He,d)$^{12}$C  &  TALYS  &  10.4634  &  $^{11}$B(p,$\gamma$)$^{12}$C  &  NACREII  &  15.9569\\
$^{11}$B(p,$\alpha$)$^{4}$He  &  NACREII  &  8.6821  &  $^{11}$B($\alpha$,n)$^{14}$N  &  NACREII  &  0.1581\\
$^{12}$B(p,$\alpha$)$^{9}$Be  &  TALYS  &  6.8854  &  $^{12}$B($\alpha,\gamma$)$^{16}$N  &  TALYS  &  10.1101\\
$^{12}$B(p,n)$^{12}$C  &  TALYS  &  12.5866  &  $^{12}$B($\alpha$,n)$^{15}$N  &  TALYS  &  7.6211\\
$^{12}$B(d,$\gamma$)$^{14}$C  &  TALYS  &  23.4847  &  $^{12}$B(d,n)$^{13}$C  &  TALYS  &  15.3083\\
$^{12}$B(d,p)$^{13}$B  &  TALYS  &  2.6535  &  $^{12}$B(d,$\alpha$)$^{10}$Be  &  TALYS  &  11.4730\\
$^{12}$B($^3$He,$\gamma$)$^{15}$N  &  TALYS  &  28.1987  &  $^{12}$B($^3$He,n)$^{14}$N  &  TALYS  &  17.3654\\
$^{12}$B($^3$He,p)$^{14}$C  &  TALYS  &  17.9913  &  $^{12}$B($^3$He,$\alpha$)$^{11}$B  &  TALYS  &  17.2073\\
$^{12}$B(n,$\gamma$)$^{13}$B  &  TALYS  &  4.8780  &  $^{12}$B(p,$\gamma$)$^{13}$C  &  TALYS  &  17.5329\\
$^{12}$B(t,$\gamma$)$^{15}$C  &  TALYS  &  18.4456  &  $^{12}$B(t,n)$^{14}$C  &  TALYS  &  17.2275\\
$^{12}$B(t,$\alpha$)$^{11}$Be  &  TALYS  &  5.7198  &  $^{9}$C($\alpha$,p)$^{12}$N  &  Wie89  &  6.7083\\
$^{9}$C($\alpha,\gamma$)$^{13}$O  &  TALYS  &  8.2234  &  $^{9}$C(d,p)$^{10}$C  &  TALYS  &  19.0586\\
$^{9}$C(n,$\gamma$)$^{10}$C  &  TALYS  &  21.2831  &  $^{9}$C(t,$\gamma$)$^{12}$N  &  TALYS  &  26.5222\\
$^{9}$C(t,p)$^{11}$C  &  TALYS  &  25.9210  &  $^{9}$C(t,$\alpha$)$^{8}$B  &  TALYS  &  18.5139\\
$^{9}$C(n,$^3$He)$^{7}$Be  &  TALYS  &  6.2806  &  $^{9}$C(n,d)$^{8}$B  &  TALYS  &  0.9246\\
$^{11}$C(p,$\gamma$)$^{12}$N  &  Tan03  &  0.6012  &  $^{11}$C(n,$\gamma$)$^{12}$C  &  Rau94  &  18.7216\\
{\bf $^{11}$C(n,$\alpha$)$^{4}$He}  &  {\bf Rau94}  &  11.4469  &  $^{11}$C($\alpha$,p)$^{14}$N  &  NACRE  &  2.9228\\
$^{11}$C(d,$\gamma$)$^{13}$N  &  TALYS  &  18.4405  &  {\bf $^{11}$C(d,p)$^{12}$C}  &  {\bf TALYS}  &  16.4971\\
$^{11}$C($^3$He,$\gamma$)$^{14}$O  &  TALYS  &  17.5742  &  $^{11}$C($^3$He,p)$^{13}$N  &  TALYS  &  12.9471\\
$^{11}$C($^3$He,$\alpha$)$^{10}$C  &  TALYS  &  7.4579  &  $^{11}$C(t,$\gamma$)$^{14}$N  &  TALYS  &  22.7367\\
$^{11}$C(t,n)$^{13}$N  &  TALYS  &  12.1833  &  $^{11}$C(t,p)$^{13}$C  &  TALYS  &  15.1861\\
$^{11}$C(t,$\alpha$)$^{10}$B  &  TALYS  &  11.1245  &  $^{11}$C($\alpha,\gamma$)$^{15}$O  &  TALYS  &  10.2196\\
$^{11}$C(t,d)$^{12}$C  &  TALYS  &  12.4644  &  $^{11}$C(t,$^3$He)$^{11}$B  &  TALYS  &  2.0010\\
$^{12}$C($\alpha,\gamma$)$^{16}$O  &  NACREII  &  7.1619  &  $^{12}$C(d,$\gamma$)$^{14}$N  &  TALYS  &  10.2723\\
$^{12}$C(d,p)$^{13}$C  &  TALYS  &  2.7217  &  $^{12}$C($^3$He,$\gamma$)$^{15}$O  &  TALYS  &  12.0756\\
$^{12}$C($^3$He,p)$^{14}$N  &  TALYS  &  4.7788  &  $^{12}$C($^3$He,$\alpha$)$^{11}$C  &  TALYS  &  1.8560\\
$^{12}$C(n,$\gamma$)$^{13}$C  &  TALYS  &  4.9463  &  $^{12}$C(p,$\gamma$)$^{13}$N  &  NACREII  &  1.9435\\
$^{12}$C(t,$\gamma$)$^{15}$N  &  TALYS  &  14.8484  &  $^{12}$C(t,n)$^{14}$N  &  TALYS  &  4.0151\\
$^{12}$C(t,p)$^{14}$C  &  TALYS  &  4.6409  &  $^{12}$C(t,$\alpha$)$^{11}$B  &  TALYS  &  3.8570\\
$^{13}$C($\alpha,\gamma$)$^{17}$O  &  TALYS  &  6.3587  &  $^{13}$C(d,$\gamma$)$^{15}$N  &  TALYS  &  16.1593\\
$^{13}$C(d,n)$^{14}$N  &  TALYS  &  5.3260  &  $^{13}$C(d,p)$^{14}$C  &  TALYS  &  5.9519\\
$^{13}$C(d,$\alpha$)$^{11}$B  &  TALYS  &  5.1679  &  $^{13}$C($^3$He,$\gamma$)$^{16}$O  &  TALYS  &  22.7932\\
$^{13}$C($^3$He,n)$^{15}$O  &  TALYS  &  7.1293  &  $^{13}$C($^3$He,p)$^{15}$N  &  TALYS  &  10.6658\\
$^{13}$C($^3$He,$\alpha$)$^{12}$C  &  TALYS  &  15.6313  &  $^{13}$C(n,$\gamma$)$^{14}$C  &  TALYS  &  8.1764\\
$^{13}$C(p,$\gamma$)$^{14}$N  &  NACREII  &  7.5506  &  $^{13}$C(t,$\gamma$)$^{16}$N  &  TALYS  &  12.3911\\
$^{13}$C(t,n)$^{15}$N  &  TALYS  &  9.9021  &  $^{13}$C(t,p)$^{15}$C  &  TALYS  &  0.9127\\
$^{13}$C(t,$\alpha$)$^{12}$B  &  TALYS  &  2.2810  &  $^{13}$C($\alpha$,n)$^{16}$O  &  NACREII  &  2.2156\\
$^{14}$C(d,n)$^{15}$N  &  Kaw91  &  7.9829  &  $^{14}$C(n,$\gamma$)$^{15}$C  &  Kaw91  &  1.2181\\
$^{14}$C($\alpha,\gamma$)$^{18}$O  &  ILCCF10  &  6.2263  &  $^{14}$C(p,$\gamma$)$^{15}$N  &  ILCCF10  &  10.2074\\
$^{14}$C(d,$\gamma$)$^{16}$N  &  TALYS  &  10.4719  &  $^{14}$C(d,$\alpha$)$^{12}$B  &  TALYS  &  0.3618\\
$^{14}$C($^3$He,$\gamma$)$^{17}$O  &  TALYS  &  18.7599  &  $^{14}$C($^3$He,n)$^{16}$O  &  TALYS  &  14.6168\\
$^{14}$C($^3$He,p)$^{16}$N  &  TALYS  &  4.9784  &  $^{14}$C($^3$He,$\alpha$)$^{13}$C  &  TALYS  &  12.4012\\
$^{14}$C(t,$\gamma$)$^{17}$N  &  TALYS  &  10.0987  &  $^{14}$C(t,n)$^{16}$N  &  TALYS  &  4.2147\\
$^{15}$C($\alpha,\gamma$)$^{19}$O  &  TALYS  &  8.9631  &  $^{15}$C($\alpha$,n)$^{18}$O  &  TALYS  &  5.0082\\
$^{15}$C(n,$\gamma$)$^{16}$C  &  TALYS  &  4.2504  &  $^{15}$C(p,$\gamma$)$^{16}$N  &  TALYS  &  11.4784\\
$^{15}$C(p,n)$^{15}$N  &  TALYS  &  8.9893  &  $^{15}$C(p,$\alpha$)$^{12}$B  &  TALYS  &  1.3683\\
$^{15}$C(p,d)$^{14}$C  &  TALYS  &  1.0065  &  $^{12}$N(n,p)$^{12}$C  &  TALYS  &  18.1204\\
$^{12}$N($\alpha$,p)$^{15}$O  &  TALYS  &  9.6184  &  $^{12}$N(n,$\gamma$)$^{13}$N  &  TALYS  &  20.0639\\
$^{12}$N(p,$\gamma$)$^{13}$O  &  TALYS  &  1.5151  &  $^{12}$N(n,d)$^{11}$C  &  TALYS  &  1.6234\\
$^{13}$N(n,$\gamma$)$^{14}$N  &  TALYS  &  10.5534  &  $^{13}$N($\alpha,\gamma$)$^{17}$F  &  TALYS  &  5.8187\\
$^{13}$N(n,p)$^{13}$C  &  TALYS  &  3.0028  &  $^{13}$N(p,$\gamma$)$^{14}$O  &  NACREII  &  4.6271\\
$^{13}$N(n,d)$^{12}$C  &  TALYS  &  0.2811  &  $^{14}$N(n,p)$^{14}$C  &  CF88  &  0.6259\\
$^{14}$N($\alpha,\gamma$)$^{18}$F  &  ILCCF10  &  4.4146  &  $^{14}$N(n,$\gamma$)$^{15}$N  &  TALYS  &  10.8333\\
$^{14}$N(p,$\gamma$)$^{15}$O  &  NACREII  &  7.2968  &  $^{15}$N($\alpha,\gamma$)$^{19}$F  &  ILCCF10  &  4.0137\\
$^{15}$N(n,$\gamma$)$^{16}$N  &  TALYS  &  2.4891  &  $^{15}$N(p,$\gamma$)$^{16}$O  &  NACREII  &  12.1274\\
$^{15}$N(p,$\alpha$)$^{12}$C  &  NACREII  &  4.9655  &  $^{14}$O(n,p)$^{14}$N  &  CF88  &  5.9263\\
$^{14}$O($\alpha,\gamma$)$^{18}$Ne  &  Wie87  &  5.1151  &  $^{14}$O($\alpha$,p)$^{17}$F  &  Bar97C  &  1.1916\\
$^{14}$O(n,$\gamma$)$^{15}$O  &  TALYS  &  13.2231  &  $^{14}$O(n,$\alpha$)$^{11}$C  &  TALYS  &  3.0035\\
$^{14}$O(n,$^3$He)$^{12}$C  &  TALYS  &  1.1475  &  $^{15}$O($\alpha,\gamma$)$^{19}$Ne  &  ILCCF10  &  3.5291\\
$^{15}$O(n,$\gamma$)$^{16}$O  &  TALYS  &  15.6639  &  $^{15}$O(n,p)$^{15}$N  &  TALYS  &  3.5365\\
$^{15}$O(n,$\alpha$)$^{12}$C  &  TALYS  &  8.5020  &  $^{16}$O(n,$\gamma$)$^{17}$O  &  Iga95  &  4.1431\\
$^{16}$O(p,$\alpha$)$^{13}$N  &  CF88  &  -5.2184\&  $^{16}$O(p,$\gamma$)$^{17}$F  &  ILCCF10  &  0.6003\\
$^{16}$O($\alpha,\gamma$)$^{20}$Ne  &  ILCCF10  &  4.7298  &  $^{17}$O(n,$\alpha$)$^{14}$C  &  Koe91  &  1.8177\\
$^{17}$O(n,$\gamma$)$^{18}$O  &  TALYS  &  8.0440  &  $^{17}$O(p,$\gamma$)$^{18}$F  &  ILCCF10  &  5.6065\\
$^{17}$O(p,$\alpha$)$^{14}$N  &  ILCCF10  &  1.1918  &  $^{17}$O($\alpha,\gamma$)$^{21}$Ne  &  CF88  &  7.3479\\
$^{17}$O($\alpha$,n)$^{20}$Ne  &  NACRE  &  0.5867  &  $^{18}$O(n,$\gamma$)$^{19}$O  &  TALYS  &  3.9549\\
$^{18}$O(p,$\gamma$)$^{19}$F  &  ILCCF10  &  7.9949  &  $^{18}$O($\alpha,\gamma$)$^{22}$Ne  &  ILCCF10  &  9.6681\\
$^{18}$O(p,$\alpha$)$^{15}$N  &  ILCCF10  &  3.9811  &  $^{19}$O($\alpha,\gamma$)$^{23}$Ne  &  TALYS  &  10.9139\\
$^{19}$O($\alpha$,n)$^{22}$Ne  &  TALYS  &  5.7132  &  $^{19}$O(n,$\gamma$)$^{20}$O  &  TALYS  &  7.6087\\
$^{19}$O(p,$\gamma$)$^{20}$F  &  TALYS  &  10.6413  &  $^{19}$O(p,n)$^{19}$F  &  TALYS  &  4.0399\\
$^{19}$O(p,$\alpha$)$^{16}$N  &  TALYS  &  2.5153  &  $^{17}$F(n,$\alpha$)$^{14}$N  &  NACRE  &  4.7347\\
$^{17}$F(n,p)$^{17}$O  &  TALYS  &  3.5429  &  $^{17}$F(p,$\gamma$)$^{18}$Ne  &  ILCCF10  &  3.9235\\
$^{17}$F($\alpha,\gamma$)$^{21}$Na  &  TALYS  &  6.5608  &  $^{17}$F($\alpha$,p)$^{20}$Ne  &  TALYS  &  4.1296\\
$^{17}$F(n,$\gamma$)$^{18}$F  &  TALYS  &  9.1493  &  $^{18}$F(n,$\alpha$)$^{15}$N  &  CF88  &  6.4187\\
$^{18}$F(n,p)$^{18}$O  &  TALYS  &  2.4375  &  $^{18}$F(p,$\gamma$)$^{19}$Ne  &  ILCCF10  &  6.4112\\
$^{18}$F(p,$\alpha$)$^{15}$O  &  ILCCF10  &  2.8822  &  $^{18}$F($\alpha,\gamma$)$^{22}$Na  &  TALYS  &  8.4810\\
$^{18}$F($\alpha$,p)$^{21}$Ne  &  TALYS  &  1.7414  &  $^{18}$F(n,$\gamma$)$^{19}$F  &  TALYS  &  10.4324\\
$^{19}$F($\alpha,\gamma$)$^{23}$Na  &  TALYS  &  10.4674  &  $^{19}$F($\alpha$,p)$^{22}$Ne  &  TALYS  &  1.6733\\
$^{19}$F(n,$\gamma$)$^{20}$F  &  TALYS  &  6.6013  &  $^{19}$F(p,$\gamma$)$^{20}$Ne  &  NACRE  &  12.8435\\
$^{19}$F(p,$\alpha$)$^{16}$O  &  NACRE  &  8.1137  &  $^{18}$Ne(n,$\alpha$)$^{15}$O  &  TALYS  &  8.1080\\
$^{18}$Ne(n,p)$^{18}$F  &  TALYS  &  5.2258  &  $^{19}$Ne(n,p)$^{19}$F  &  TALYS  &  4.0212\\
$^{19}$Ne(p,$\gamma$)$^{20}$Na  &  ILCCF10  &  2.1924  &  $^{19}$Ne(n,$\alpha$)$^{16}$O  &  TALYS  &  12.1348\\
$^{20}$Na(n,$\alpha$)$^{17}$F  &  TALYS  &  10.5427  &  

\enddata

References: NACRE~\citep{NACRE}, NACRE~II~\citep{NACRE2,NACRE2b}, DAACV04~\citep{Des04}, ILCCF10~\citep{ILCCF10b},
CF88~\citep{CF88}, MF89~\citep{MF89}, Boy93~\citep{Boy93}, Bal95~\citep{Bal95}, Hei98~\citep{Hei98}, Rau94~\citep{Rau94},
Des99~\citep{Des99}, Bea01~\citep{Bea01}, Tan03~\citep{Tan03}, Wan91~\citep{Wan91}, Efr96~\citep{Efr96}, Wie87~\citep{Wie87},
Bar97~\citep{Bar97}, Koe91~\citep{Koe91}, Cam08~\citep{Cam08}, And06~\citep{And06}, Ser04~\citep{Ser04}, Wag69~\citep{Wag69},
Has09~\citep{Has09}, Wie89~\citep{Wie89}, FK90~\citep{FK90}, Bru91~\citep{Bru91}, Bec92~\citep{Bec92}, Iga95~\citep{Iga95},
Cyb08~\citep{Cyb08}, Miz00~\citep{Miz00}. \\ 
Reaction and references shown in {\bf bold face} : re-evaluated rate for the improved network, see text.

\label{t:network}
\end{deluxetable}

\end{document}